# Observation of Majorana Fermions in Ferromagnetic Atomic Chains on a Superconductor

Stevan Nadj-Perge*[1], Ilya K. Drozdov*[1], Jian Li*[1], Hua Chen*[2], Sangjun Jeon[1], Jungpil Seo[1], Allan H. MacDonald[2], B. Andrei Bernevig[1] and Ali Yazdani[1]†

[1]Joseph Henry Laboratories & Department of Physics, Princeton University, Princeton, NJ 08544, USA.

[2]Department of Physics, University of Texas at Austin, Austin, TX 78712, USA.

* These authors contributed equally to this work

† To whom correspondence should be addressed. Email: yazdani@princeton.edu

Abstract

Majorana fermions are predicted to localize at the edge of a topological superconductor, a state of matter that can form when a ferromagnetic system is placed in proximity to a conventional superconductor with strong spin-orbit interaction. With the goal of realizing a one-dimensional topological superconductor, we have fabricated ferromagnetic iron (Fe) atomic chains on the surface of superconducting lead (Pb). Using high-resolution spectroscopic imaging techniques, we show that the onset of superconductivity, which gaps the electronic density of states in the bulk of the Fe chains, is accompanied by the appearance of zero energy end states. This spatially resolved signature provides strong evidence, corroborated by other observations, for the formation of a topological phase and edge-bound Majorana fermions in our atomic chains.

Topological superconductors are a distinct form of matter that is predicted to host boundary Majorana fermions (*1-3*). These quasi-particles are the emergent condensed matter analogs of the putative elementary spin-1/2 particles originally proposed by Ettore Majorana (*4*) with the intriguing property of being their own anti-particles. Super-symmetric theories in particle physics and some models for dark matter in cosmology motivate an ongoing search for free Majorana particles (*5, 6*). The search for Majorana quasi-particle (MQP) bound states in condensed matter systems is motivated in part by their potential use as topological qubits to perform fault-tolerant computation aided by their non-Abelian characteristics (*7, 8*). Spatially separated pairs of MQP pairs can be used to encode information in a nonlocal fashion, making them more immune to quantum decoherence. Early proposals for the detection of MQPs were based on the properties of superfluid $^3$He, on exotic fractional quantum Hall states, or on correlated superconductors (*9-12*). The focus in the last few years has shifted to the search for these exotic fermions in weakly interacting synthetic systems in which proximity to a conventional Bardeen-Cooper-Schrieffer (BCS) superconductor is used in concert with other electronic properties to create the topological phase that hosts MQPs.

The idea that MQPs can be engineered in the laboratory grew from the theoretical observation that proximity induced superconductivity on the surface state of a topological insulator is topological in nature (*13*). Pairing on a "spin-less" Fermi surface (*1*), created in this case by the spin-momentum locking of topological surface states, must be effectively p-wave to satisfy the pair-wavefunction anti-symmetry requirement and is therefore topological. This approach was later extended to systems in which a semiconductor nanowire with strong spin orbit interactions in a parallel magnetic field is in contact with a superconductor (*14, 15*). Experimental efforts to implement the nanowire proposal have uncovered evidence for a zero bias peak (ZBP) in tunneling spectroscopy studies of hybrid superconductor-semiconductor nanowire devices, as expected in the presence of the MQP states of a topological superconductor (*16-19*). However, the ZBPs detected in such devices could also be caused by the Kondo effect or disorder (*20-24*). A key disadvantage of the nanowire studies is that they lack the ability to spatially resolve ZBP features in order to demonstrate that they are localized at the boundary of a gapped superconducting phase. Here we introduce a method of fabricating one-dimensional topological superconductors and detecting their MQPs that achieves both spatial and spectral resolution.

*Magnetic Atomic Chains as a Platform for Topological Superconductivity*

Magnetic atom chains on the surface of a conventional s-wave superconductor have been proposed to provide a versatile platform for the realization of topological superconductors (*25*). This platform lends itself to the detection of MQPs using the spectroscopic imagining techniques of scanning tunneling microscopy (STM). In the absence of intrinsic spin-orbit coupling, (*25*) and related theoretical work (*26-30*) showed that a topological phase emerges in an atomic chain when its magnetic atoms have a spatially modulated spin arrangement, for example a spin helix. The spin texture of the chains emulates the combination of spin-orbit and Zeeman interactions required to create a topological phase. Helical spin configurations are however much less common in atomic chains than simple ferromagnetic and antiferromagnetic ones or may be more influenced by disorder (*31*). We therefore explore an alternate, more realizable scenario by placing a Fe chain on the surface of Pb (Fig. 1A). We will show that the essential ingredients for topological superconductivity in this scenario are the ferromagnetic interaction between Fe atoms realized at the Fe-Fe bond distance and the strong spin-orbit interaction in superconducting Pb (*32*). Our approach is related to earlier proposals for topological superconductivity using half-metal ferromagnets or metallic chains placed in contact with superconductors in the presence of spin-orbit interactions (*33-35*).

To illustrate the key ingredient of our approach, we first consider an idealized ferromagnetic chain of Fe atoms described by a tight-binding model calculation (Section 1 of (*36*)). We use hopping parameters appropriate for d orbitals of bulk Fe to compute the band structure of a freely suspended linear Fe chain (Fig. 1B). The large exchange interaction results in a fully occupied majority spin band with the Fermi level ($E_F$) residing in the minority spin bands. Coupling the Fe chain to a BCS superconductor with strong spin-orbit interaction (such as Pb), we find that the spin-orbit interactions lift many of the degeneracies in the chain's band structure shown in Fig. 1B, while at the same time allowing for the occurrence of p-wave superconductivity (Section 1 of (*36*)). Since only the Fe *d*-bands will be strongly spin-polarized, other bands are unlikely to influence the topological character of the system, whether they reside mainly on the Fe chains or on the substrate. Remarkably, for large exchange interaction, topological superconductivity is ubiquitous to the type of band structure shown in Fig. 1B—occurring for nearly all values of the chemical potential (Fig. 1C, (*36*) for details). In this idealized situation, the number of minority spin bands which cross the Fermi level is almost always odd, making the presence of MQPs at the ends of the chains almost guaranteed.

We consider another idealized situation for topological superconductivity by modeling a ferromagnetic chain embedded in a 2D superconductor, which allows us identify its signatures in STM measurements (Section 2 of (*36*)). The spatially resolved density of states (DOS) of this 2D model at positions on the chain differs from that of a BCS superconductor by the presence of Yu-Shiba-Rusinov in-gap states (Fig. 1, C and D) (*37-41*). These calculations also exhibit the spatial and spectroscopic signatures of MQPs at the chain ends (Fig. 1, D and E). Other more realistic models for our experimental system are also worth exploring (see below) and non-topological phases can occur for some chain geometries. These model studies of proximity-induced superconductivity on Fe chains demonstrate that topological states can be identified using STM by establishing: 1) ferromagnetism on the chain, 2) spin-orbit coupling in the host superconductor (or at its surface) 3) a superconducting gap in the bulk of the chain, and finally 4) a localized ZBP due to MQPs at the ends of the chain. One can over constrain these conditions by providing evidence that the system has an odd number of band crossings at $E_F$. The disappearance of edge-localized ZBPs when the underlying superconductivity is suppressed provides an additional check to show that the MQP signature is associated with superconductivity and not with other phenomena, such as the Kondo effect (*20-22*).

*Ferromagnetic Fe Atomic Chains on the Pb(110) Surface*

To fabricate an atomic chain system on the surface of a superconductor with strong spin-orbit coupling, we used a Pb (110) single crystal, which we prepare with cycles of in situ sputtering and annealing. Following sub-monolayer evaporation of Fe on the Pb surface at room temperature and light annealing, STM images (temperature was 1.4K for all experiments reported here) show large atomically ordered regions of the Pb (110) surface, as well as islands and chains of Fe atoms that have nucleated on the surface (Fig. 2A). The islands appear to provide the seed from which chains self-assemble following the anisotropic structure of the underlying surface. Depending on growth conditions, we find Fe chains as long as 500 Å, usually with an Fe island in the middle (inset, Fig. 2A). In longer chains, the ends are separated from the islands in the middle by atomically ordered regions that are 200 Å long. High-resolution STM images show that the chains (with an apparent height of about 2 Å) are centered between the atomic rows of Pb (110), display weak atomic corrugation (5-10 pm) and strain the underlying substrate (Fig. 2, B-D). Approximate periodicities of 4.2Å and 21Å measured on the chain show that the Fe chain has a structure that is incommensurate with that of the underlying Pb surface. To identify the atomic structure of our chains, we performed density functional theory (DFT) calculations of Fe on the Pb (110) surface which show that strong Fe-Pb bonding results in a partially submerged zigzag chain of Fe atoms between Pb (110) atom rows (Fig. 2, E and F; see Section 3 of (*36*) for DFT details). From these calculations, we find that among several candidate structures with the experimental periodicity, a three-layer Fe zigzag chain partially submerged in Pb has the lowest energy and gives contours of constant electron density most consistent with our STM images.

      We use a combination of spectroscopic and spin-polarized measurements to demonstrate that Fe atomic chains on Pb (110) satisfy the criteria (conditions 1-4 above) required to demonstrate a one-dimensional topological superconductor. First, we discuss spin-polarized STM studies that show experimental evidence for ferromagnetism on the Fe chains and strong spin-orbit coupling on the Pb surface (Fig. 3, A-C). Using Cr STM tips, which have been prepared using controlled indentation of the tip into Fe islands, we measured tunneling conductance (dI/dV) at a low bias (V = 30 mV) as a function of magnetic field perpendicular to the surface on both the chains and on the Pb substrate (Fig. 3, A and B). Our preparation of spin-polarized tips were validated by also performing experiments on Co on Cu (111), now a standard system (*40*) for verifying

spin-polarized STM capabilities, in situ during the same experimental runs. The spin-polarized measurements on the Fe chains show hysteresis loops characteristic of tunneling between two ferromagnets with the field switching only one of them (at around 0.25 T) (*42, 43*). The strength of the spin-polarized STM signal varies along the chain, likely due to the electronic and structural properties of our zigzag Fe chains. We find that tunneling with the same tip on the Pb (110) surface far from the Fe chains also shows field-dependent conductance. In contrast to the asymmetric behavior observed on the chains, the field dependence on the substrate is symmetric with field, as expected for tunneling into non-magnetic Pb, but still shows the switching behavior that is due to magnetization reversal of the tip. Similar tunneling magnetoresistance curves have been previously reported for tunneling from a ferromagnet into semiconductors and have been attributed to spin-polarized tunneling in the presence of strong spin-orbit interactions (*44*). The field-dependent signal on Pb is consistent with a preference for spins to be in the plane of the surface, in which case further polarization of the tip's magnetization perpendicular to the surface suppresses tunneling conductance. The size of this signal depends on the orientation of the tip's magnetization relative to that of the spins at the surface. This observation is consistent with a Rashba-like ($\mathbf{k} \cdot \sigma$) spin-orbit coupling at the Pb (110) surface upon which our ferromagnetic Fe chain is self-assembled.

Our DFT calculations confirm that the zigzag Fe chains in Pb (110) are likely ferromagnetic (Section 3 of (*36*)), as expected given that the distance between the Fe atoms is close to that of bulk Fe. These calculations also demonstrate that Fe chains on Pb have exchange energy *J* of about 2.4 eV and a magnetic anisotropy energy (1.4 meV/Fe atom) that is large enough to suppress thermal fluctuations in the magnetic configuration. The large anisotropy also implies that very large fields are required to switch the chain's magnetization orientation, consistent with our finding that low fields only reorient the tip's magnetization.

Our model studies demonstrate that topological superconductivity occurs when an odd number of 1D bands cross the Fermi energy. We now focus on obtaining information on the electronic structure of our Fe chains from STM spectroscopy. As shown in Fig. 3D, the Fe chain contribution to the DOS has two distinct peaks separated by about 0.9 eV, one above and the other below $E_F$. To understand the origin of this double-peak structure, we have carried out a number of model calculations, including those based on tight binding and DFT for our specific zigzag Fe structures identified by comparing DFT to STM images (Fig. 2, C-F). The results of these calculations, for both suspended and Pb-imbedded chains (Sections 3 and 4 in (*36*)) indicate that $E_F$ for such a system lies in the Fe minority bands and that the double-peak structure is related to structure-specific features of the minority-band bonding pattern. Moreover, the tight-binding model that reproduces the experimentally observed double-peak structure (Fig. 3E) shows the number of band crossing at $E_F$ to be odd (phase diagram in Section 4 in (*36*)). Although momentum-resolved DOS measurements with the STM will be required to more directly probe the band structure of the chains, the correspondence between model calculations and the measured DOS is very suggestive that the band structure of these chains is consistent with topological superconductivity. All the chains we examined in search for MQPs share the above structural and electronic characteristics.

*Superconductivity on Fe Atomic Chains and Signatures of MQP*

We use high resolution spectroscopy and spectroscopic mapping, with both normal and superconducting tips, to establish the signature of both the pairing gap on our Fe chains and of the MQP at their ends at low temperatures (1.4 K) (Figs. 4 and 5). In Fig. 4, we show spatially resolved spectra, as well as spectroscopic maps with a normal tip, on both the substrate and the Fe chain. Examining the spectra away from the Fe chain on the Pb surface, we resolved a superconducting gap that can be modeled with the thermally broadened (T=1.4 K) BCS density of states with Pb's bulk pairing gap ($\Delta_s =$ 1.36 meV) (Fig. 4A, lowest curve). In the middle of the chains, spatially resolved spectroscopic measurements show the formation of features in the Pb gap, with asymmetric edges at roughly ±1 meV, which evolve to include a sharp ZBP that is prominently detected about 15-20 Å from the ends of the chain. Figure 4, B-E show such ZBPs at two ends of the same chain that grow out of an Fe island. Approaching the island, the spectra on both sides of the chain also show enhanced low-energy DOS, but at these locations, the structure of the chains is not known (example in Figure S14 in (*36*)).

Spectroscopic maps in Fig. 4F visualize the spatial structure of excitations at different energies and clearly show the localized nature of the ZBPs at one end of the chain and the delocalized nature of the excitations at higher energies throughout the chain. These maps also resolve the spatially modulated decay of the ZBP away from the chain ends that is anticipated from model calculations (Section 2 of (*36*)). Our ability to correlate the location of the ZBP with the end of our atomic chains is one of the main experimental requirements for interpreting that this feature is associated with the predicted MQP of a topological superconductor. We have confirmed the robust observation of ZBPs at the end of multiple chains (more than ten), measured with different tips, on different experimental runs with freshly prepared Pb and Fe chains (see Section 5 of (*36*) for more examples).

To obtain more precise information about the pairing gap induced on the Fe chains, we have also carried out spectroscopic measurements with superconducting tips, prepared in situ by contacting a W tip with the Pb surface (Fig. 5). Away from the Fe chains, the spectra measured with these tips show the behavior expected for tunneling between two superconductors, with sharp peaks at energies corresponding to the sum of the BCS gaps on the tip $\Delta_T$ and the substrate $\Delta_S$ (Fig. 5A). The singularity in DOS at the gap edge in this case allows us to resolve finer features within the thermally broadened DOS of our chains. At the end of the chains, spectroscopy with the superconducting tips shows a sharp peak at $|eV| = \Delta_T$ corresponding to the peaks at these locations identified as ZBPs when measured with the normal tips (Fig. 5B). Spectroscopic maps with the superconducting tip at this energy also show this feature to be the most pronounced at the end of the chains (Fig. 5D, middle panel) as expected for bound MQPs. At other locations in the middle (bulk) of the chains (Fig. 5, C and D), this signature of MQPs is greatly suppressed, but there are sharp peaks at higher energies $|eV| > \Delta_T$ that can be attributed to the edge of a pairing gap at about 200-300 μeV. Using a model calculation of scattering of electrons in an Fe chain from a Pb(110) surface, we estimate that the effective Rashba spin-orbit interaction strength induced on the chain by the Pb substrate (Section 6 of (*36*)) is about 0.11 eV-Å. Using this value in our simple linear chain calculation of Fig. 1 (B and C) yields a value for the p-wave gap of around 100 µeV, which is qualitatively consistent with our experimental results (Section 1 of (*36*)). Separate calculations for an embedded zigzag structure using Pb's bulk atomic spin-orbit coupling also yield a similar value (Section 4 of (*36*)). The qualitative agreement between the theoretical and experimental estimates of p-wave gap is remarkable considering the many uncertainties that we have not taken into account, e.g., surface relaxation, surface dipole, lattice mismatch, and modification of s-wave pair potentials near the surface of Pb. The size of this pairing gap is also consistent with the low-energy background tunneling conductance (50% of the normal state) observed at 1.4 K in the middle of the wire with the normal tip, as shown in Fig. 4 (see also Section 9 of (*36*) for more details). We also note that our experiments are in a regime where coupling strength between the STM tip and the MQP at the end of our chains is small in energy, as compared to the thermal energy, resulting in the suppression of the zero bias conductance ($1.3 \times 10^{-4}$ $e^2/h$, as shown in Fig. 4) from the ideal value of $2e^2/h$ (*45*). Evidence for a gap in the middle of our chains, with a value that is in line with model calculations, together with ZBPs at their ends completes the list of experimental observations required to conclude that superconductivity on our ferromagnetic atomic chains is topological in nature.

Finally, we comment on the particle-hole asymmetric features at energies above the minimum gap of our chain measured with normal tips (near ±1 meV) throughout the chain, shown in Fig. 4). These features in the spectra are reproducible and reminiscent of the band of Yu-Shiba-Rusinov in-gap states found in the 2D model calculation of Fig. 1D. Our DFT calculations show that there is significant charge transfer from the Pb substrate to the Fe chain, resulting in a linear dipole moment density (per Å) of 0.02 e·Å perpendicular to the Pb surface. As in the case of isolated magnetic impurities, this potential makes tunneling in to and out of the in-gap states asymmetric (see also results of simulations in Fig. 1D that include such a potential) (*40, 41*). These in-gap states also change approaching the end of our chains; however, they are always distinct in energy from the ZBP we associate with the MQP, even at the end of the chain (see fig. S14 in (*36*) for example).

*Experimental Checks for Interpretations other than MQPs*

To address alternatives to the MQP interpretation of the ZBP, we have carried out a number of control experiments. First, we examine the possibility of the ZBP being due to a Kondo resonance, which has been raised in the context of semiconductor nanowire experiments where there is experimental evidence for a Kondo effect (*21, 22*). To address this issue, we have carried out experiments in weak magnetic fields (0.1T) that suppress superconductivity in the Pb substrate. All features associated with the gap in the middle of the chains and the ZBP at their ends disappear in the absence of superconductivity in Pb (Section 7 of (*36*)). If the ZBPs at the ends of the chains were due to the Kondo effect, we expect that increasing the DOS near $E_F$ in the normal state would only enhance the ZBP rather than suppress it. The importance of superconductivity to the formation of the ZBP also rules out alternative scenarios in which such peaks appear due to disorder effects. Second, our spin-polarized STM measurements do not show significant magnetization change at the end of the chain, which discounts the unlikely scenario in which the pairing gap at the end of the chain is strongly suppressed and therefore gives an apparent ZBP. Third, structural and potential defects in our substrate cannot produce in-gap states (including ZBPs) in accordance with Anderson's theorem and previous experiments, which show that, for conventional s-wave superconductors, in-gap states are not induced by non-magnetic adsorbates or step edges (*40, 41*).

Finally, we add that we have found very short Fe chains (~ 30-40 Å) in which the ZBPs are strongly suppressed (Section 8 of (*36*)). This observation suggests that coupling the end states to each other suppresses the signatures we are associating with MQPs. Although more detailed length dependence experiments are needed to characterize the decay length of MQP end states more quantitatively, this experiment establishes that our ZBPs are not associated with disorder at the end of our chains. More specifically, our model calculations show the wavefunction of the MQP at one end of our hybrid chain-superconductor system to have a combination of power-law decay (on the Fermi length scale) and an exponential decay (related to the p-wave pairing's coherence length) as a function of distance from the end of the chains (Section 2 of (*36*)) and Ref. (*46*)). Our experimentally observed decay of the ZBP over 15-20 Å (Figs. 4 and 5) is likely associated with the power-law decay and the effective Fermi wavelength in our chain and is consistent with the suppression of the MQP signatures in chains of double that length. The small ZBP splitting that may be present even in our longest chains is smaller than our energy resolution, which is about 100 μeV for measurements with the superconducting tip. Besides the coupling between the MQPs on either side of a chain, there are situations where multiple channels on the chain can give rise to multiple MQP at the same end of the chain (Section 2 of (*36*)). Generically, a perturbation can couple and split these MQPs, unless they are protected by a symmetry (see for example, (*47*)) of the system, resulting in the absence of topological superconductivity. Ultimately, the splitting of our MQPs needs to be experimentally tested using higher-resolution measurements. Our ability to characterize splitting is limited by thermal broadening (1.4 K), which accounts for the width of the experimental features and contributes to the background tunneling conductance at zero bias—so-called quasi-particle poisoning of MQP (Section 9 of (*36*)) Future studies will require milli-kelvin STM measurements, a capability that has already been demonstrated in experiments on other exotic superconductors (*48*). Overall, based on the results of these control experiments together with the observation of all 4 enumerated experimental signatures, we conclude that our results are consistent with the realization of a topological superconducting state with localized MQPs.

*Outlook*

The experimental system described here demonstrates a platform for future experiments to manipulate MQPs and to realize other related one or two-dimensional topological superconducting phases. An obvious extension of our experiments is to 2D islands of ferromagnetic films on the surface of Pb. Provided these films are thin enough, a few monolayer as in our chains, pairing could be stabilized at a reasonable temperature and the edges of these islands could harbor propagating Majorana modes. The detailed structure of such modes, whether they can be chiral or fully in-gap, depends on the spin-orbit coupling configuration. Searching for other systems with both even or odd number of band crossing at $E_F$ on Pb can be used to further test the concept behind our studies and should show both topological and non-topological superconducting phases. Although the phase with the even number of crossing at the $E_F$ is not topological, it is a model system to form the Fulde-Ferrell-Larkin-Ovchinnikov (FFLO) phase that has a modulated gap structure with a periodicity related to (in the simplest case) the difference between the two Fermi points (1D) or circles (2D) (*49, 50*). STM spectroscopic mapping can be used to characterize the modulated gap structure of this system and provide evidence for an FFLO phase.

Ultimately, manipulation of MQPs is required to demonstrate braiding and the non-Abelian characteristics of these quasi-particles. We have recently proposed that application of a parallel magnetic field to ring like magnetic atomic structures fabricated on a thin film superconductor can be used to generate edges between trivial and topological regions with MQP in such rings (*51*). The rotation of the parallel magnetic field (that does not perturb superconductivity in thin films) can then be used to manipulate and braid MQPs in ring like atomic structures. This proposal applies to both the spin-helix arrangement of magnetism as well as to the ferromagnetic chains studied here. In addition to manipulating MQPs, parallel field applied to a chain on a thin film superconductor can also be used to drive the chains between topological and trivial superconducting phases. Ultimately, atomic manipulation techniques with the STM can be used to fabricate complex magnetic structures in which MQP may be engineered and manipulated.

# References and Notes

**Acknowledgments:**
We thank F. von Oppen, L. Glazman, D. Loss, A. Stern, J. Alicea, M. Franz, R. Lutchyn, C-K. Shih, B. Feldman, M. Randeria, P. Lee, and N. P. Ong, for helpful discussions. The work at Princeton has been primarily supported by Majorana Basic Research Challenge grant through ONR-N00014-14-1-0330, ONR-N00014-11-1-0635, ONR- N00014-13-10661, NSF-MRSEC programs through the Princeton Center for Complex Materials DMR-0819860, and NSF CAREER DMR-095242. Work at UT Austin was supported by ONR-N00014-14-1-0330 and by the Welch Foundation grant TBF1473. This project was also made possible using the facilities at Princeton Nanoscale Microscopy Laboratory supported by grants through NSF-DMR-1104612, ARO-W911NF-1-0262, ARO-MURI program W911NF-12-1-0461, DOE-BES, DARPA-SPWAR Meso program N6601-11-1-4110, LPS and ARO-W911NF-1-0606, and the W. Keck Foundation. AY, AB, and Princeton University have filled a provisional patent related to this work. S. N-P acknowledges support of the EU Marie Curie program (IOF 302937). JL acknowledges support of Swiss National Science Foundation. AY acknowledges the hospitality of the Aspen Center for Physics, supported under NSF Grant No. PHYS-1066293.

**Figure Captions:**

**Fig. 1. Topological superconductivity and Majorana fermions in Ferromagnetic atomic chains on a superconductor.** (**A**) Schematic of the proposal for Majorana quasi-particle (MQP) realization and detection: a ferromagnetic atomic chain is placed on the surface of strongly spin-orbit (SO) coupled superconductor and studied using scanning tunneling microscopy (STM) (**B**) Band structure of a linear suspended Fe chain before introducing spin-orbit coupling or superconductivity. The majority spin-up (red) and minority spin-down (blue) d-bands labeled by azimuthal angular momentum m are split by the exchange interaction $J$ (degeneracy each band is noted by the number of arrows). (**C**) Regimes for trivial and topological superconducting phases are identified for the

band structure shown in (**B**) as a function of exchange interaction in presence of SO coupling. The value *J* for Fe chains based on DFT calculations is noted (Section 1 and 3 of (*36*)). (**D**) Model calculation of the local density of states (LDOS) of the atomic chain embedded in a 2D superconductor (Section 2 of (*36*)). Left panel shows an image of the chain and the locations at which the LDOS is represented in the right panel; the LDOS curves are offset for clarity. In-gap (Shiba) and zero-energy (MQP) features in LDOS are noted. (**F**) Spatially resolved LDOS calculated at various energies noted at the bottom using the same model. Red (blue) indicates regions of the high (low) LDOS.

**Fig. 2. Atomic Fe chains on Pb (110)**. (**A**) Topograph of the Pb (110) surface after growth of Fe showing Fe islands and chains indicated by white arrows and atomically clean terraces of Pb (regions with the same color) with size exceeding 1000 Å. Lower-right inset shows the anisotropic atomic structure of the Pb(110) surface with interatomic distances in the two directions, $a = 4.95$ Å and $a/\sqrt{2} = 3.5$ Å, as expected for the face-centered cubic crystal structure. Upper-left insets show images of several atomic Fe chains and islands from which they grow (scale bar corresponds to 50 Å). (**B**) Fine scale topograph. The color-coded lines correspond to the line traces in (**C**) and (**D**). The scale bar corresponds to 20 Å. (**C**) A cross section of the image in (**B**) along the chain measured on top of the chain (light brown) and along two different lines next to the chains (purple and blue) showing atomic corrugation of the chain and strain induced in the substrates. Both are influenced by the incommensurate placement of the Fe atoms in Pb. (**D**) Apparent STM height profile on the substrate (black) and across the atomic chain (green). (**E** and **F**) Atomic structure of the zigzag chain as calculated using DFT (Section 3 of (*36*)) along (**A**) and across (**F**) the chain. The Fe chain structure that has the lowest energy in the DFT calculations matches the structural features in the STM measurements (**B-D**). All measurements were carried out at 1.4 K.

**Fig. 3. Spin-polarized measurements and normal state characterization of Ferromagnetic chains.** (**A**) Spatially average (over region of about 100 Å) STM

tunneling conductance as a function out-of-plane applied magnetic field H on the atomic chain and the substrate measured with a spin-polarized bulk Cr/Fe tip (inset shows a schematic of the measurement, set point V = 30 mV, I = 0.75 nA). Tip switching occurs at ± 0.25 T (**B**) Difference between conductance on and off the chain shown in (**A**). (**C**) Topography of the chain colorized by the conductance at ±1 T from low conductance (dark blue) to high conductance (dark red). (**D**) (Top) STM point spectra of the atomic chain and the substrate. (Bottom) DOS computed using tight binding model of a zigzag Fe chain as describe in Section 4 of (*36*).

**Fig. 4. Spectroscopic mapping of atomic chains and ZBPs.** (**A**) STM spectra measured on the atomic chain at locations corresponding to those indicated in panels (**B) and (C**). For clarity, the spectra are offset by 100 nS. The red spectrum shows the ZBP at one end of the chain. The gray trace measured on the Pb substrate can be fit using thermally broadened BCS DOS (dashed gray line, fit parameters $\Delta_s$ = 1.36 meV, T = 1.45 K). (**B, C**) Zoom-in topography of the upper end (**B**) and lower end (**C**) of the chain and corresponding locations for spectra marked (1-7). (**D, E**) Spectra measured at marked location as in (**B**) and (**C**). (**F**) Spatial and energy-resolved conductance maps of another Fe atomic chain close to its end, which shows similar features in point spectra as in (**A**). Conductance map at zero bias (middle panel) shows increased conductance close to the end of the chain. Scale bar corresponds to 10 Å. We note that the localization length of the MQP observed here is a factor of 10 or smaller in length than the distance from the end to the islands that form in the middle of the chains.

**Fig. 5. High energy-resolution spectra obtained using a superconducting tip.** (**A**) Point spectra on the superconducting substrate (with a gap of $\Delta_S$) using the superconducting tip (with a gap of $\Delta_T$) shows a peak at $\Delta_T + \Delta_S$. Upper schematic shows alignment of the BCS DOS for the tip and the sample as a function of bias, which results in the conductance at $\Delta_T + \Delta_S$ and suppressed conductance at lower voltages. (**B**) STM

spectra at the end of the Fe chain showing a peak at $\Delta_T$ which corresponds to the ZBP (inset shows schematics of the DOS alignment between the tip and the sample). The shape of the spectra and the higher peak indicate that the ZBP resides in a gap even at the end of the wire. (**C**) Point spectra in the middle of the Fe atomic chain showing a peak at $\Delta_T + 300$ µV, signaling the approximate value of the effective p-wave gap in the bulk of the chain. (**D**) Topographic image of the atomic chain and spatially resolved conductance maps for $|eV_1| = \Delta_T = 1.36$ meV and $|eV_2| = \Delta_T + 0.3$ meV. The left image indicates where the point spectra in panels (**B**) and (**C**) are taken. The scale bar corresponds to 20 Å.

**Supplementary Materials**

www.sciencemag.org

Materials and Methods

Supplementary Sections 1-9

Figs. S1-S16

Table S1

References: *(20-23, 25, 26, 30-31, 46, 52-59)*

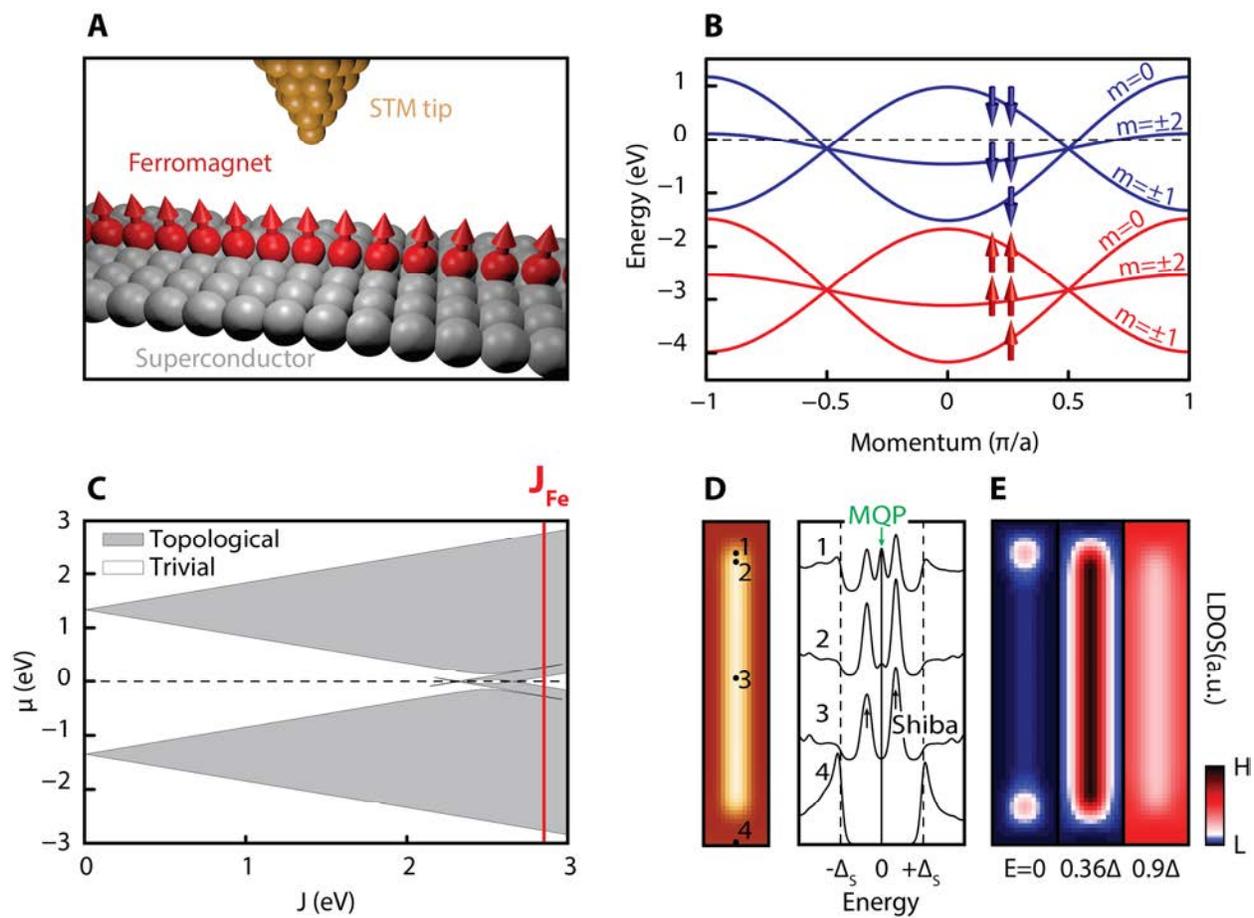

**Figure 1**

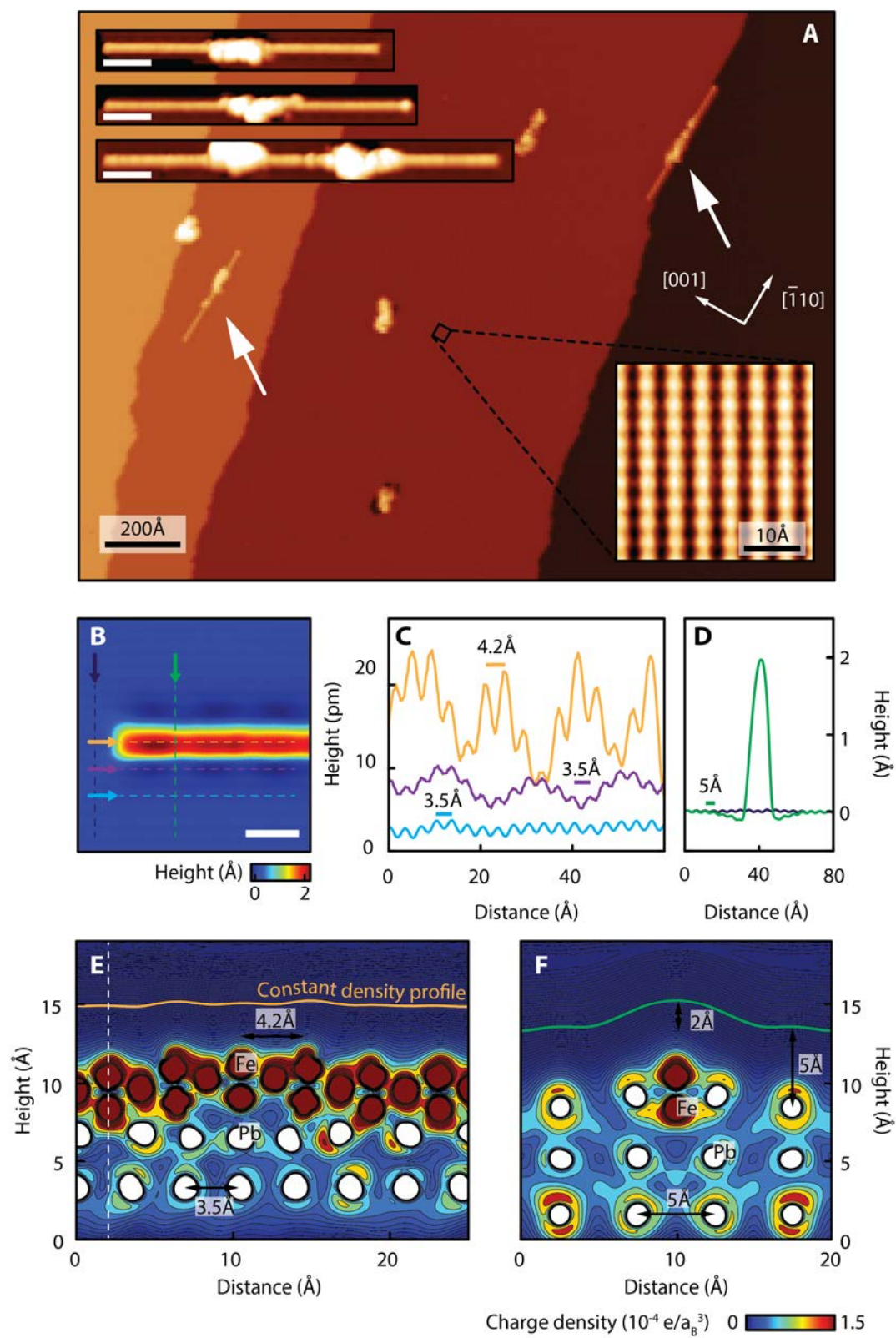

**Figure 2**

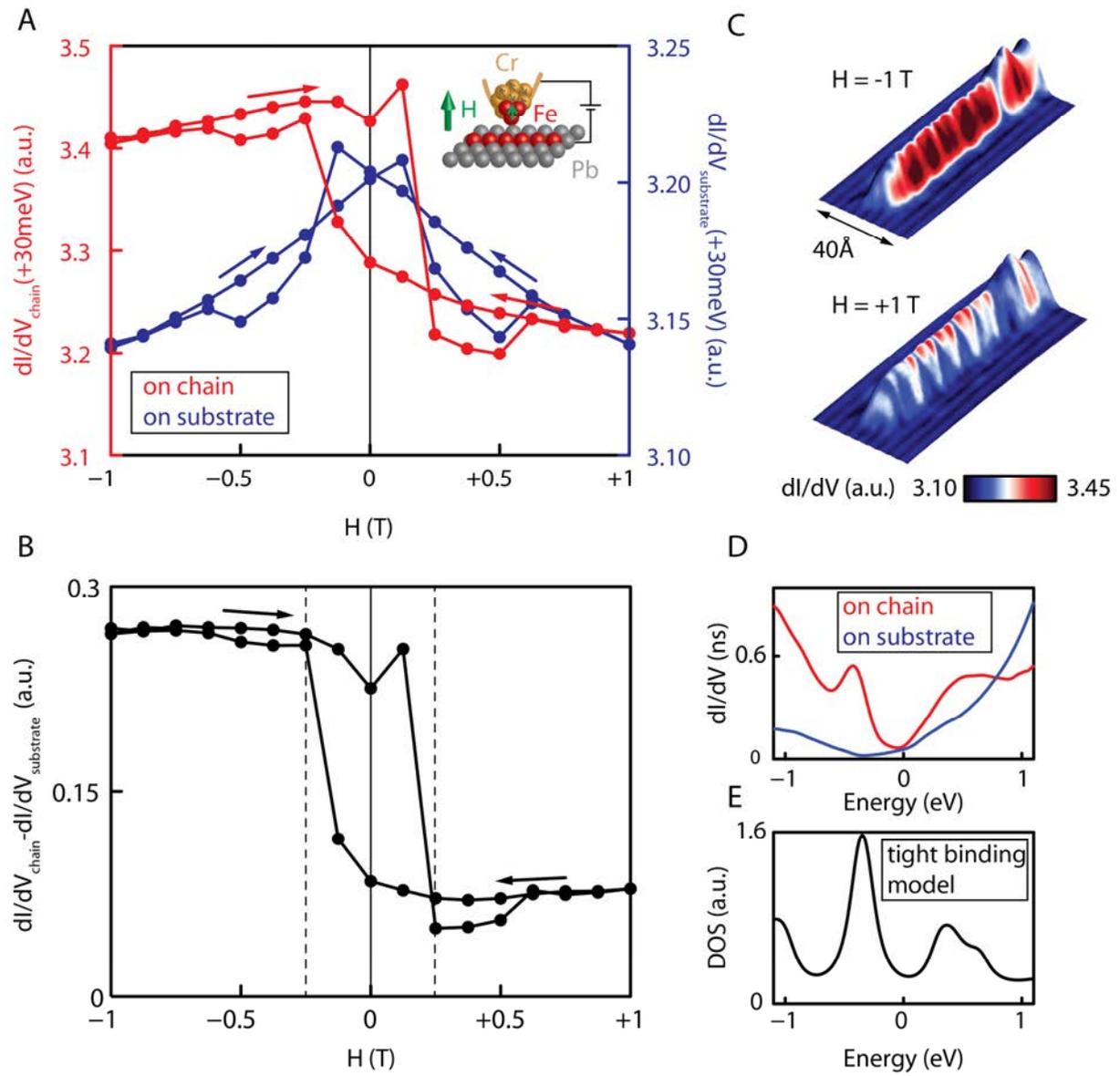

**Figure 3**

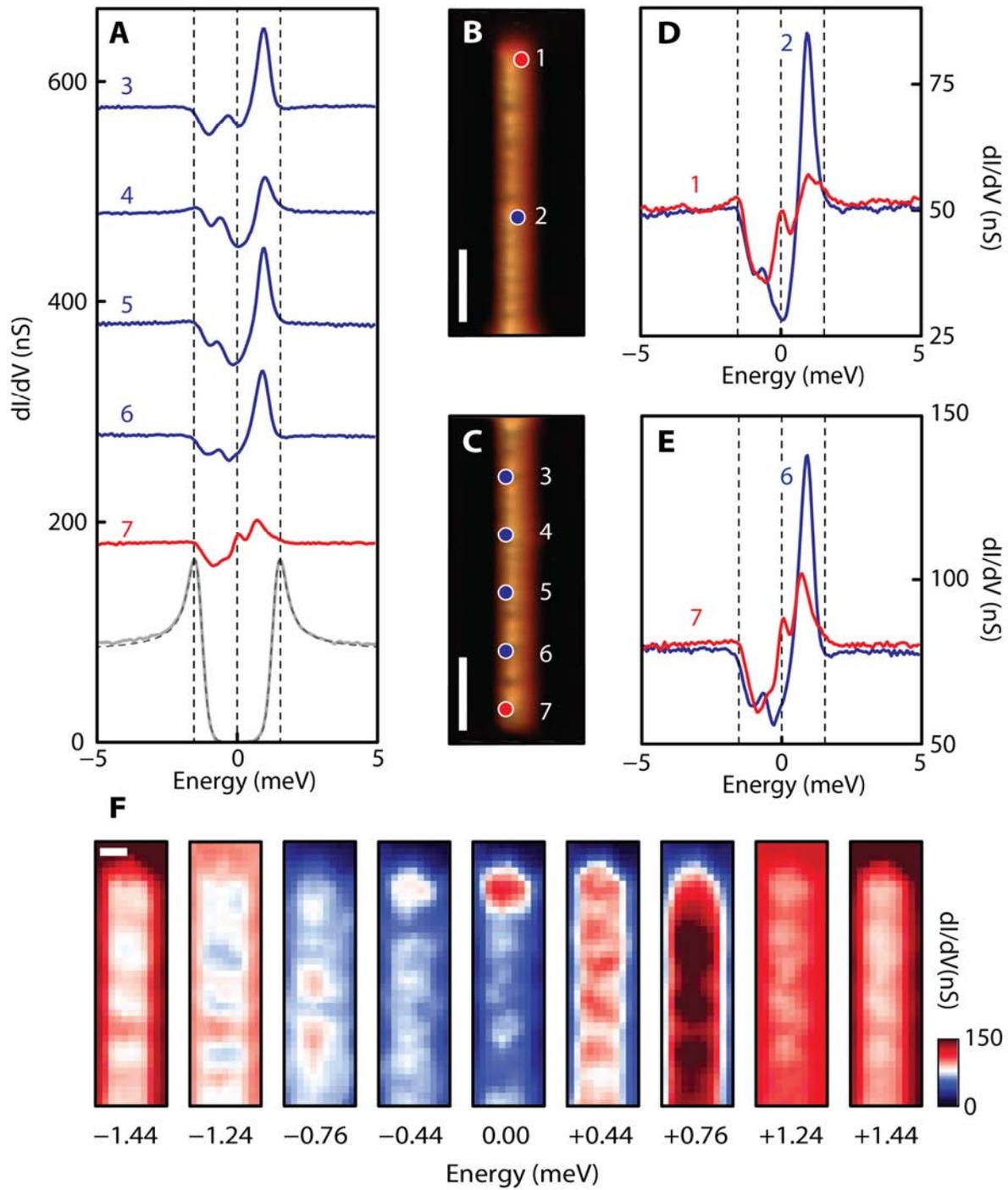

**Figure 4**

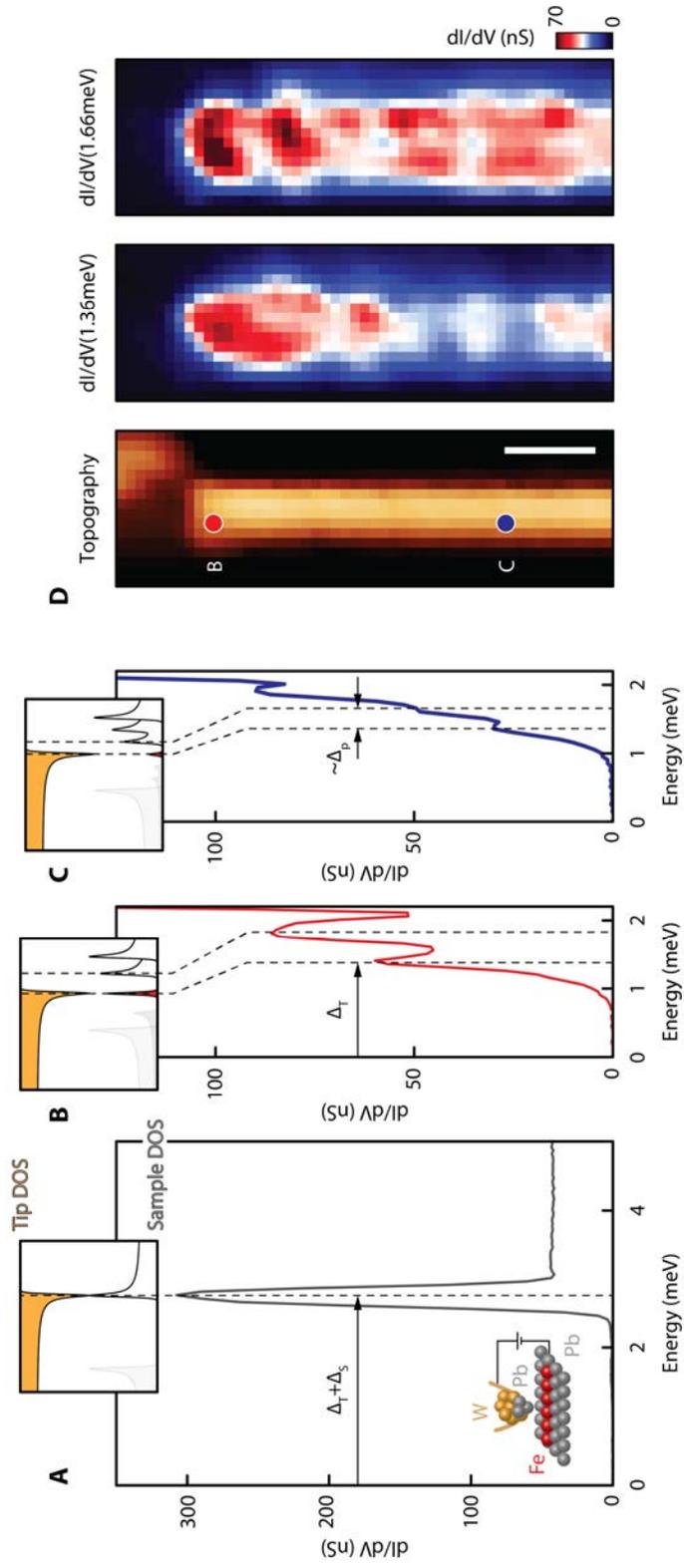

**Figure 5**

# Supplementary Materials for

## Observation of Majorana Fermions in Ferromagnetic Atomic Chains on a Superconductor

Stevan Nadj-Perge, Ilya K. Drozdov, Jian Li, Hua Chen, Sangjun Jeon, Jungpil Seo, Allan H. MacDonald, B. Andrei Bernevig, and Ali Yazdani

correspondence to: yazdani@princeton.edu

**This PDF file includes:**

Materials and Methods
Supplementary Text: Sections 1-9
Figs. S1 to S16
Tables S1

## Materials and Methods
Experimental setup and Data Acquisition

The STM measurements were carried out on a homebuilt ultra-high vacuum scanning tunneling microscope operating at T=1.4 K, which is capable of storing two samples simultaneously. The Pb (110) single crystal surface was prepared by repeated cycles of Argon ion sputtering (E= 1keV for 15 minutes) and annealing (at T = 250 °C for 40 minutes). Atomic chains were synthesized by 0.1-0.5 monolayer deposition of Fe on the Pb surface at room temperature and light post-annealing (T = 175°C for 6-8 minutes).

Non-spin-polarized data is taken with an etched tungsten (W) tip. Spectroscopy of the Pb(110) superconducting gap and in-gap states of the atomic chains were measured with a standard lock-in technique using 40 μV modulation and set-point bias in the range of 5-8 mV, well above the value corresponding to the superconducting gap (1.37mV). For the measurement using a superconducting tip reported in Fig. 5, we have repeatedly indented the Pb sample with the W tip until a large enough Pb cluster was transferred onto the tip apex. The resulting tip had a fully developed superconducting gap which was verified both on normal Cu (111) and superconducting Pb (110) surfaces.

Spin-polarized measurements were performed using etched bulk chromium (Cr) tips. The spin polarization of the tip was first checked on the system of cobalt islands on Cu (111). Afterwards, the same macroscopic tip was conditioned in situ on Fe clusters and then used to image Fe chains on Pb. In cases where the spin contrast was lost, it would be restored by controllable indentation of the tip into Fe clusters. The spin contrast of the resulting tip was verified by observing the difference of conductance on chains at B= ±1T. Typically, 30 mV bias, 750pA setpoint current, and 5 mV lock-in modulation were used.

## Supplementary Text
### Section 1: Linear Fe Chain Band Structure and Topological Superconductivity

In this section, we consider a suspended linear chain of Fe atoms to illustrate the key ingredients of our approach for the realization of topological superconductivity and MQPs (see also main Fig. 1 B and C). We consider only the 3$d$ orbitals of a straight Fe chain, described by the following mean-field tight-binding Hamiltonian:

$$H_{MF}= H_{SK}+ H_{J}+ H_{SO}+ H_{R} \quad (S1)$$

The terms in above equation are discussed below:

$$H_{SK} = \sum_{\langle ij \rangle \alpha' \alpha \sigma} t_{\alpha'\alpha} c^{\dagger}_{i\alpha'\sigma} c_{j\alpha\sigma} \quad (S2)$$

which is the Slater-Koster approximation tight-binding Hamiltonian, where $i$ and $j$ label sites, $\langle ij \rangle$ implies a restriction to nearest neighbor sites, $\sigma$ labels spin, and $\alpha' \alpha$ label the five d orbitals on each site. $t_{\alpha'\alpha}$ are real Slater-Koster integrals that that are, in general, dependent on the direction cosines of the vector connecting nearest neighbors for each orbital pair.

In the case of a linear chain d-orbitals where the different azimuthal angular moment are decoupled, and $V_{dd\delta}$, $V_{dd\pi}$, and $V_{dd\sigma}$ are the hopping parameters for m=0, |m|=1, and

|m|=2 bands respectively, we adopt the Slater-Koster parameter values from (*52*), which are listed in Table S1. The Stoner-theory spin splitting term is:

$$H_J = -J\hat{m} \cdot s \tag{S3}$$

where the typical size of *J* in Fe is 2-3 eV depending on the dimensionality of the system. *J* is larger in lower dimensional systems because the spin polarization is larger. The enhanced spin polarization is in turn due to narrower d-bands and a consequent reduction in its band energy cost. The direction of the spin polarization $\hat{m}$ is taken to be perpendicular to the chain, the easy-direction predicted by the DFT presented in Section 3 below. We also have on-site atomic spin-orbit coupling term, which for the 3*d* orbitals of Fe is not very large ($\lambda_{SO}$ ~60 meV):

$$H_{so} = \lambda_{so} L \cdot s \tag{S4}$$

A key ingredient is the orbital-independent Rashba spin-orbit coupling due to the hybridization between the Fe *d* orbitals and the *p* orbitals of the Pb substrate:

$$H_R = it_R \sum_{\langle ij \rangle \gamma \tau} c_{i\gamma}^\dagger c_{j\tau} (\hat{d}_{ij} \times \sigma_{\gamma\tau}) \cdot \hat{z} \tag{S5}$$

Here $\hat{d}_{ij}$ is a unit vector pointing from site *i* to site *j*, $\hat{z}$ is the same direction as the spin polarization $\hat{m}$, and $\gamma$, $\tau$ are spin indices. We take $t_R$=0.05 eV as estimated in Section 6 below.

To account for the proximity-induced superconductivity in the Fe chain we assume the following local, orbital-independent, s-wave pairing potential

$$H_{pair} = 2\Delta \sum_\alpha \left( c_{\alpha\uparrow}^\dagger c_{\alpha\downarrow}^\dagger + c_{\alpha\downarrow} c_{\alpha\uparrow} \right) \tag{S6}$$

where we have chosen a real Slater-Koster basis for the *d* orbitals and set the pairing Δ to be 1.5 meV, the same as the *s*-wave gap in Pb. For these parameters, we find that this system has an odd number of crossings (Fig. 1C main text), with the Fermi level ($E_F$) in a spin minority band. Solving the Bogoliubov-de Gennes (BdG) equations, we found a *p*-wave gap of ~0.1 meV at $E_F$ (Fig S1B), in qualitative agreement with the gap estimated from the STM measurements with the superconducting tip as described in the main text.

To calculate the Majorana number, which allows us to identify the topological trivial and non-trivial regions for our model system (Fig 1 C of the main text), we follow previous work (*53*) and note that the BdG Hamiltonian can be written in the Majorana basis $\gamma_k \equiv (\gamma_{k,a}, \gamma_{k,b})^T$ as

$$H_{BdG} = \frac{i}{2} \sum_k \gamma_k^\dagger \tilde{A}_k \gamma_k \tag{S7}$$

where $\tilde{A}_k$ is an antisymmetric matrix. $\tilde{A}_k$ can be obtained from the BdG Hamiltonian written in Nambu basis through a unitary transformation *U*

$$\tilde{A}_k = -iU H_{BdG}(k) U^\dagger, \quad U = \frac{1}{\sqrt{2}} \begin{pmatrix} 1 & 1 \\ -i & i \end{pmatrix}. \tag{S8}$$

The Majorana number of an infinite 1D system is then calculated as

$$\mathcal{M} = \text{sgn}\left[\text{Pf}(\tilde{A}_{k=0})\text{Pf}(\tilde{A}_{k=\frac{\pi}{a}})\right], \quad (S9)$$

where Pf stands for Pfaffian of an antisymmetric matrix. When $\mathcal{M} = -1$ the chain is topologically nontrivial and isolated MQPs should appear in a long chain. The results of such calculations as a function of chemical potential and *J* are shown in Fig. 1 C.

**Section 2: Atomic chains embedded in a 2D superconductor: Real Space Signature of MQP & Multi-MQPs in Chains**

To consider how a realistic STM experiment would probe topological superconductivity in our atomic chains, we consider the situation of a ferromagnetic chain embedded in a 2D lattice, which we view as an approximation to the surface of a bulk superconductor (see Fig.S2). This model is analogous to other similar models recently investigated in the literature, where the atomic magnetic chain induces a band of hybridized so-called Yu-Shiba-Rusinov states within the host superconductor's gap (*25, 26, 30*) The Hamiltonian for this 2D model is given by

$$\begin{aligned}
H_{2D} = & \sum_{n_x,n_y} \left\{ (4t-\mu)\mathbf{c}^\dagger_{n_x,n_y}\mathbf{c}_{n_x,n_y} + (\Delta c^\dagger_{n_x,n_y,\uparrow}c^\dagger_{n_x,n_y,\downarrow} + h.c.) \right. \\
& -[\mathbf{c}^\dagger_{n_x,n_y}(t\sigma_0 + i\alpha\sigma_2)\mathbf{c}_{n_x+1,n_y} + \mathbf{c}^\dagger_{n_x,n_y}(t\sigma_0 - i\alpha\sigma_1)\mathbf{c}_{n_x,n_y+1} + h.c.] \right\} \\
& + \sum_{n_x,n_y}{}' \mathbf{c}^\dagger_{n_x,n_y}((J/2)\sigma_3 + U\sigma_0)\mathbf{c}_{n_x,n_y},
\end{aligned} \quad (S10)$$

where $\mathbf{c}_{nx,ny}{}^\dagger = (c_{nx,ny,\uparrow}{}^\dagger, c_{nx,ny,\downarrow}{}^\dagger)$ are the electron creation operators, σ's are the Pauli matrices, *μ* is the chemical potential, *t* is the hopping strength, *α* is the spin-orbit coupled hopping; the second summation $\sum'_{nx,ny}$ is limited to the sites that are in the immediate vicinity of magnetic adatoms (highlighted in red in Fig. S2); *J* stands for the exchange coupling energy between the magnetic adatoms (treated as classical spins) and the superconductor host atoms; *U* stands for the local scalar potential induced by the magnetic atoms in the chain. Since the purpose of this section is conceptual, we will use dimensionless parameters *t*=1, $\Delta = \alpha = 0.3$ and *U*=0 throughout this section. As shown in Fig. S3, this model can be investigated numerically to explore the behavior of a system with multiple bands, as in the case of a double chain we discuss below.

Figure S3 shows the results of our numerical simulation of the model for a finite-size one-atom wide ferromagnetic chain (Fig S2a) as function of parameters *μ* and *J*. The phase diagram for this system (Fig. S3a) shows topological trivial regions (Blue) and large nontrivial regions at larger values of *J* in a parameter range relevant to our Fe chains. This 2D model can be used to calculate the spatial structure of MPQs as shown in Figure S3d, and the spatial and energy-resolved tunneling density of state (Figure 1D & 1E in the main text and Figure S3d).

*Multiband situation*

We can extend the 2D model to study a scenario with multiple relevant bands at $E_F$, by considering a coupled chain illustrated schematically in Fig. S3b. Figure S4 shows the results of our simulation for the case of two coupled ferromagnetic chains as a function of $J$ or $\mu$. Again, there can generically be topologically nontrivial phases, characterized by the presence of zero-energy MQPs. However, in general coupled chains there can be in general more than one pair of MQP end states (for example, there are two pairs of zero energy end states in-between the two topologically nontrivial regions in Fig. S4). In principal, generic perturbations would split multiple MQPs, unless protected by some symmetry (see such a case in Ref. (*47*)). In the absence of such protection, the even number of the MQP are split and the system is not topological. Just as in the case of coupling between MQP from two ends of the same chain, which would split a single MQP at each end, the coupling between multiple MQP on the same end can only be experimentally verified (see discussion in Section 9 for current experimental resolution).

*MQP wavefunction in atomic chains*

The 2D models examined above provide information on the spatial decay of MQP wavefunctions in the realistic situation when our 1D system is embedded in a superconducting host. To understand the influence of the host environment we tune the coupling $t'$ between the chain and the host (Fig. S2a) and examine the spatial decay of MQPs when this coupling is weak or strong. Fig. S5a shows the evolution of the Majorana wavefunctions when $t'$ is tuned gradually from 0 to its normal value of 1. One noticeable effect is that the Majorana wavefunction is increasingly localized at the ends when $t'$ is increased. Despite the fact that this enhanced localization is in part due to a shorter p-wave coherence length (which is responsible for the exponential decay of the wavefunction), a closer inspection of the wavefunctions in logarithm scale in Fig. S5b shows that the envelope of the wavefunction becomes highly nonlinear (in logarithm scale) close to the ends of the chain. This nonlinearity can be well fitted by an additional power-law prefactor $1/\sqrt{x}$. Related theoretical efforts in the physics of semiconductor nanowires coupled to superconductors have also found the importance of power-law decay for this higher dimensional situation (*46*).

## **Section 3: Details of density functional theory calculations**

Density functional theory (DFT) calculations were performed using projector augmented wave (PAW) pseudopotentials as implemented in Vienna Ab-initio Simulation Package (VASP) (*54, 55*). The energy cutoff for the plane wave functions was set to 300 eV, and all structures were relaxed until the force on each atom was smaller than 0.01 eV/Å. The Pb(110) substrate was modeled by a $6 \times 4$ surface unit cell with 5 layers of Pb atoms. The Pb atoms in the lower two layers were fixed in their respective bulk positions while the other 3 layers wer allowed to relax. A vacuum layer of about 14 Å was used to decouple the 2D slabs in neighboring supercells along the surface normal. A $4 \times 4 \times 1$ Monkhorst-Pack k-point mesh (*56*) was used for structural relaxation and total energy calculations.

We first identify stable sites for a single Fe atom on a Pb(110) surface. Starting from several different initial configurations, we found that there are only two stable sites for Fe on Pb(110) after ionic relaxation. These are illustrated in Figs. S6A and S6B. The most stable site is in the trench between parallel lines of Pb atoms on the top surface.

This theoretical finding is consistent with the position of the Fe chain observed in experiment, as shown in Fig. 2D in the main text. In both cases shown in Fig. S6, the Fe atom is submerged below the Pb surface because of the strong p-d bonding between Pb and Fe and because the Fe atomic radius is much smaller than that of Pb. In STM images of Fig. 2 in main text, the chains appear 2 Å above the Pb surface with a corrugation of 4.3Å along the chain. Since the period of the Pb(110) surface along the chain direction is 3.5Å, a reasonable choice of the size of the unit cell for the Fe-chain/Pb(110) composite structure is 6 times of the Pb(110) surface unit cell along the chain direction, which gives a mismatch of 4.3×5-3.5×6=0.5 Å, or 2.4%. We then consider three possible Fe chain structures consistent with this periodicity, as shown in Fig S6 C to E. The three structures correspond to chains with 1, 2, and 3 layers of Fe atoms in the Pb(110) trenches. We did not consider alloyed chains made of both Fe and Pb atoms, because formation of well defined chains on the terraces of Pb(110) after annealing suggests that alloying with Fe is not energetically favorable on the Pb surface. Because of the strong bonding between Pb and Fe and the small atomic radius of Fe, chains with 1 and 2 layers of Fe atoms (Fig. 1 C and D, respectively) have almost the same height as the neighboring Pb atoms which is inconsistent with the 2 Å height found experimentally. After ionic relaxation these two structures also do not give the correct 4.3 Å period between neighboring Fe atoms in the top layer. In contrast, the Fe chain with three layers of Fe atoms (Fig. S6 E) gives the correct height of 2 Å, and the correct separation of 4.3 Å between two neighboring Fe atoms in the top layer. This structure also has lower formation energy per Fe atom compared to the two other structures. We therefore conclude that the three-layer zigzag Fe chain structure is the most likely geometry of the experimentally observed chains.

The local magnetic moments on the Fe atoms in a Fe chain on the Pb(110) substrate are found to range from 2.03 $\mu_B$ to 2.77 $\mu_B$ depending on position. The value of *J* was estimated to 2.4eV, which is similar to that used in calculation of band structure from various tight binding models described in the main text and the supplementary sections. The magneto-crystalline anisotropy, which is 1.4 meV per Fe with the easy axis along the surface normal, was calculated by evaluating the total energy difference, including spin-orbit coupling, between configurations with all moments along the surface normal and along the chain direction. The simulations used to match the STM image at $E_F$ were obtained by integrating the ground state charge density in a 50 meV window right below the Fermi energy (Figure 2E and 2F in the main text).

We also performed DFT calculations of the DOS for the geometry of zigzag triple Fe chains embedded in un-relaxed Pb (110) crystal, the results of which are shown in Fig. S7. This calculation can be compared with the tight binding calculation discussed in the next section both of which show similar double peak structure and relative depression of DOS near EF as seen experimentally (Fig. 2 D main text).

**<u>Section 4: Tight Binding Model of the Zigzag Fe Chain</u>**

We have also used a tight binding model of the three layer zigzag Fe chain to understand the STM measurements of DOS, and its implications for the number of band crossing at $E_F$ in this system. The theoretical approach here is the same as that in Section 1, where we used a tight-binding Hamiltonian for Fe's 3d orbital as in equation (S1) above. The only difference is that we exclude the substrate Rashba term used in that section for spin-orbit coupling at the surface (see below though). The Slater-Koster parameters used in the calculation are also listed in Table S1. All other parameters, such as exchange and the strength of spin-orbit coupling on Fe are also the same as those in Section 1. The structure of the chain is that shown in Figure S8, with Fe's bonded at 90° angle, which is similar to the chain structure we have determined based on comparison of STM data to DFT calculations (Section 3). In Figure S9, we show the tight-binding band structure and its DOS for this triple chain structure that shows remarkable similarity to experimental results in Fig.3D in the main text. In addition to the double peak structure seen experimentally, our tight binding model also shows the number of band crossing within about 0.5eV of $E_F$ to be odd—a condition required for the formation of topological superconducting phase and MQPs. As in Section 1, we calculate the Majorana number as function of chemical potential and *J* to construct a phase diagram for this system (Fig. S10).

      We have also extended this tight binding calculation by embedding a zigzag Fe chain in a Pb(110) structure (also modeled by a tight binding model (*57*) with stronger atomic spin orbit interaction, $\lambda_{SO}$=0.6eV). We have examined the properties of Fe-Pb system for different hybridization of the zigzag chain with the lattice. We have also repeated the tight binding calculation for isolated chain in several other scenarios, by including terms beyond nearest neighbors (not included in the equation S2 above), or by adding different potentials to second or third rows of Fe atoms in a chain to simulate the influence of charge transfer from Pb. Overall, we find that for a wide range of hybridization with Pb or considering these other details, the double peak structure in Fig. S9 survives but it does get broadened like the DFT calculation shown in Fig. S7. Finally, we note that adding a 1.2meV pairing gap only on the Pb states in hybridized zigzag-chain-Pb system, we find, a p-wave gap of about 0.1meV, very similar to found for a linear chain in Section 1. The qualitative agreement between the theoretical and experimental estimates of the p-wave gap is remarkable considering the many uncertainties that were not taken into account, e.g., surface relaxation, surface dipole, lattice mismatch, and s-wave pairing potential near the Pb surface.

## Section 5: More examples of ZBP at the ends of atomic chains

    We have observed the ZBP and its localized structure at the end of our Fe chains on many different chains prepared on multiple experimental runs measured with different tips. Figure S11 shows energy and spatially resolved STM conductance maps that directly visualize the ZBP in multiple chains. Additional spectra are shown in Figure S12.

## Section 6: Estimate of Rashba spin-orbit coupling on atomic chains

    The strength of spin-orbit coupling plays a key role in determining the size of the induced p-wave gap in our atomic chains. Given the strong strength of atomic spin-orbit coupling in Pb (1.33 eV), as compared to that of Fe (0.06 eV), we expect that the major contribution to the spin-orbit coupling in our atomic chains is due to its coupling to the Pb substrate. The inversion symmetry breaking due to the presence of the surface allows spin-flip hopping in Pb to have a part that is antisymmetric under inversion, which results in a Rashba spin-orbit coupling.

      To estimate the size of the Rashba spin-orbit interaction on our chain, we consider a simple straight Fe chain commensurate with the Pb(110) surface unit cell along the

[$\bar{1}$10] direction. The Fe atom in the unit cell is coupled to the top layer Pb atom through $p-d$ bonding with $Vpd\sigma = -2.17 Vpd\pi$, with $Vpd\pi$ left as a parameter. $Vpd\pi$ depends on the distance between the neighboring Fe and Pb atoms and we estimate it to be on the order of 1 eV. Since we are interested in physics close to the Fermi energy we approximate the influence of the substrate by the zero energy value of the self-energy as following (*31*):

$$H_{\text{eff}} = H_{\text{Fe}} + \text{Re}[\Sigma_S(\omega = 0)] \quad \text{(S11)}$$
$$= H_{\text{Fe}} + \text{Re}[h_t^\dagger g_S(\omega = 0)]$$

where $h_t$ is the hopping matrix between the Fe chain and the Pb substrate, and $g_S$ is the surface Green's function of the Pb(110) substrate, which can be obtained through the iterative method (*58*) using previously determined (*57*) tight-binding parameters. The matrix elements of $h_t$ can generally be written as

$$\langle i | h_t^\dagger | m \, \mathbf{k}_\| \rangle = \sum_{\mathbf{r}_j} \langle i | h_t^\dagger | m \mathbf{r}_j \rangle e^{-i\mathbf{k}_\| \cdot \mathbf{r}_j} \equiv [H_t^\dagger(\mathbf{k}_\|)]_{im}, \quad \text{(S12)}$$

where $i$ labels the position, orbital, and spin of a Fe atom, $m$, $k_\|$ label the orbital and wave vector of the Pb substrate, $H_t$ is a matrix in the orbital basis of Fe and Pb, and is a function of $k_\|$. Therefore the self-energy due to the substrate is

$$\Sigma_S(k_x) = h_t^\dagger g_S h_t = \sum_{k} H_t^\dagger(\mathbf{k}_\|) G_S(\mathbf{k}_\|) H_t(\mathbf{k}_\|), \quad \text{(S13)}$$

where in order to calculate $\Sigma_S(\omega = 0, k_x)$, we need to carry out the integral over $k_y$ for each $k_x$. Note that the Pb-induced Rashba spin-orbit coupling is included through the self-energy $\Sigma_S(\omega, k_x)$.

Instead of a single parameter in a simple tight binding model, the effective Rashba spin-orbit coupling from $\Sigma_S$ will be orbital dependent and have a nontrivial structure in momentum space, which imply power law tails in real space. We define an average of the effective Rashba spin-orbit coupling at momentum $k_x$ as

$$\langle t_R \rangle(k_x) \equiv \frac{1}{2} \text{Im}\left[ \langle \tilde{\uparrow} | \Sigma_S(0, k_x) | \tilde{\downarrow} \rangle - \langle \tilde{\downarrow} | \Sigma_S(0, k_x) | \tilde{\uparrow} \rangle \right], \quad \text{(S14)}$$

With $|\tilde{\uparrow}\rangle, |\tilde{\downarrow}\rangle$ being defined as

$$|\tilde{\uparrow}\rangle \equiv \frac{1}{\sqrt{5}}(1,1,1,1,1)^T \otimes |\uparrow\rangle, |\tilde{\downarrow}\rangle \equiv \frac{1}{\sqrt{5}}(1,1,1,1,1)^T \otimes |\downarrow\rangle \quad \text{(S15)}$$

where the five dimensional vector is in Fe d-orbital basis representation. The value $\langle t_R \rangle(k_x)$ is proportional to $V_{pd\pi}^2$ as evident from the definition of $\Sigma_s$. Fig. S13 depicts the $\langle t_R \rangle$ as a function of $k_x$. One can see that it shows antisymmetry of the Rashba spin-orbit coupling under inversion. Its size can reach 0.05 $V_{pd\pi}^2$ for certain $k_x$. The slope of the curve close to $k_x = 0$ may be used to define an effective Rashba spin-orbit coupling parameter for use in less realistic models. From the slope of the curve around $k=0$ we can estimate $E_{so}$ to be around 50 meV which as discussed before gives an estimates for the p-wave gap $\Delta_p \sim 100$ μeV. We also note that the Rashba spin-orbit coupling may in addition be enhanced by the local electric fields due to the charge transfer between Pb and Fe. This is however difficult to quantify.

### Section 7: Measurements in a weak Magnetic Field: A check for the Kondo effect

ZBPs can occur due to the formation of a Kondo resonance that appears when many body interactions result in the screening of magnetic impurities in metals. Kondo ZBPs' have been shown in other systems that many be candidates for topological superconductors: in particular it has been subject of studies in semiconductor nanowires. (*20-22*). In the presence of the magnetic field, the Kondo peak first splits with the Zeeman energy before it is suppressed and typically this suppression is expected to happen when $E_z = g\mu_B B > k_B T_K$ where $E_z$ is Zeeman splitting (*g*, is Landé *g*-factor, $\mu_B$ – Bohr magneton, *B* magnetic field) and $T_K$ being Kondo temperature.

In superconductors, such shown for example in (*59*) for single magnetic impurities placed on the surface of Pb, when $T_K$ is of the order of a few Kelvin, the Kondo peak can survive magnetic fields of the order of 1T. Since Pb is a type I superconductor applying a modest field of $B = 100$mT suppress completely the pairing but the Kondo effect should exist beyond this transition. If the ZBPs reported here do not originate from MQPs but from the Kondo effect, they would persist beyond the superconductor-normal metal transition. However in our measurements, as soon as superconductivity is suppressed, all the states, including zero bias peaks at the chain's ends disappear and the spectrum becomes featureless (see Fig. S14). Since the ZBPs are observed only in the superconducting state, we conclude that our peaks are not related to Kondo physics. This experimental check not only rules out Kondo physics in our system but also makes unlikely other scenarios--such as weak (anti-) localization, (*23*)--in which superconductivity is not directly required for the formation of this ZBP.

### Section 8: Suppression of signature of MQP in short chains

The ZBP reported in Fig. 4 and Fig. 5 of the main text as well as corresponding spatial maps shown in supplementary section 5 are all obtained on chains which are at least 50 Å long (the typical lengths were in the range between 50 Å – 150 Å). In the few exceptional cases we have also found several examples of very short chains (approximately ~ 30 Å long, consisting of ~ 20 atoms). Note that these chains, although short, still have the apparent height of the regular chains and their size is larger than the size of the unit cell used in DFT calculations. For this reasons we expect that these short chains have similar structure to the longer chains considered in previous section and in the main text of the paper.

Figure S15 summarizes the measurements on these short atomic chains. In Fig. S15a and S15b, a comparison of the renormalized spatial maps for longer chain and short chains is given for two different energies. At zero bias, the end state of the longer chain is pronounced, whereas the corresponding end state in short chains is clearly suppressed. Fig. S15c and Fig. S15d show several spatial linecut maps taken along the short chains. None of them shows features at zero bias, further corroborating that ZBP are indeed suppressed in the case of short chains.

These results indicate that at the chain end we have an electronic state and rules out trivial structural related effects in which zero bias conductance peak would be a consequence of some specific configuration of the last few atoms. Note that by studying atomic chains of different lengths one may provide important information about the electronic properties of the chain's end state. Therefore more systematic studies are needed to determine the length dependence of the observed features. Nevertheless, we note that the absence of ZBP for short chain is consistent with the interpretation of the ZBP as MQPs.

### Section 9: Influence of Temperature on Background & Splitting of the ZBP

The experimental measurements reported in this study have been carried out at 1.4K, which gives rise to energy broadening that influence the energy widths and background conductance. As describe in the main text, we estimate a p-wave gap of order 200 µeV (from measurements with the superconducting tip, Fig. 5, see discussion below as well). A simple estimate based on a BCS-type DOS shows that at this temperature, we expect a background which is about 50% of the normal state, consistent with measurements with the normal tip presented in Fig. 4 of the main text.

Using the superconducting tip, with a sharp BCS coherence peak, it is however possible to obtain better resolution data and we can estimate the energy resolution of such measurements (Fig. 5 in main text) and put a limit on our current ability to resolve energy splitting of the ZBP. Without making any specific assumptions about the functional form of the peaks observed in superconducting tip spectroscopy experiment, we extract the experimental line-width from Fig 5A in the main text by comparing the width of the coherence peak to that of a Lorentzian with 250 µeV FWHM (Fig S16). We use the extracted line-width to fit the ZBP feature (peak at $E=\Delta_T$ in Fig5B and C) as well as to construct a smooth background to account for the spectral weight of thermally broadened, in-gap quasiparticles away from zero bias. This gives a crude estimate of about 20% of the signal observed at the ZBP peak position at the end of the chain (Fig 5B) and 30% in the middle (Fig 5C) to be due to thermal broadening effects. The extracted linewidth also provides an upper estimate for ZBP splitting of about 150ueV (minimal splitting between 250µeV Lorentzians that is possible to resolve) limited by the current experimental resolution.

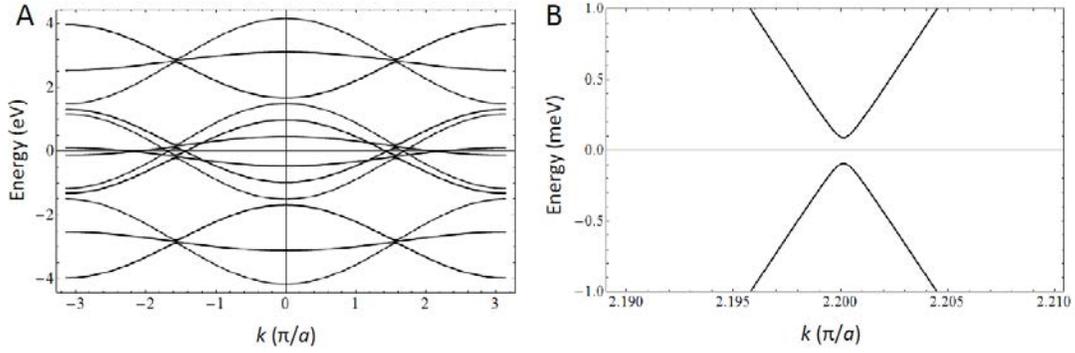

**Fig. S1.**

(A) Eigenenergies of the Bogoliubov-de Gennes Hamiltonian for the straight Fe chain model. There are in total 20 bands, some of which are nearly degenerate due to the small size of the spin-orbit coupling. (B) Spectrum of the BdG spectrum near one of the *p*-wave gaps.

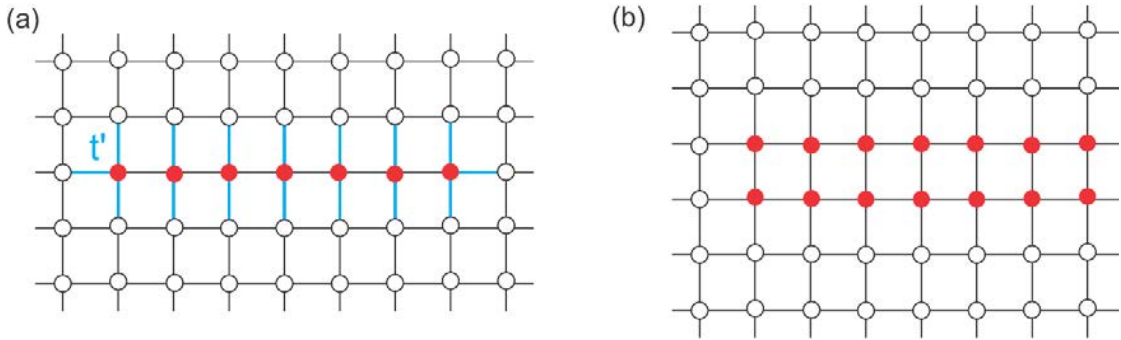

**Fig. S2**

Two-dimensional lattices with a subset of sites (one chain in (a) and two chains in (b); highlighted in red) subject to an exchange field from magnetic atoms. In panel (a), the bonds that couple the chain and the rest of the 2D lattice are highlighted in blue. Tuning this coupling ($t'$) also allows us to investigate the qualitative difference between Majorana wavefunctions in a ferromagetic chain.

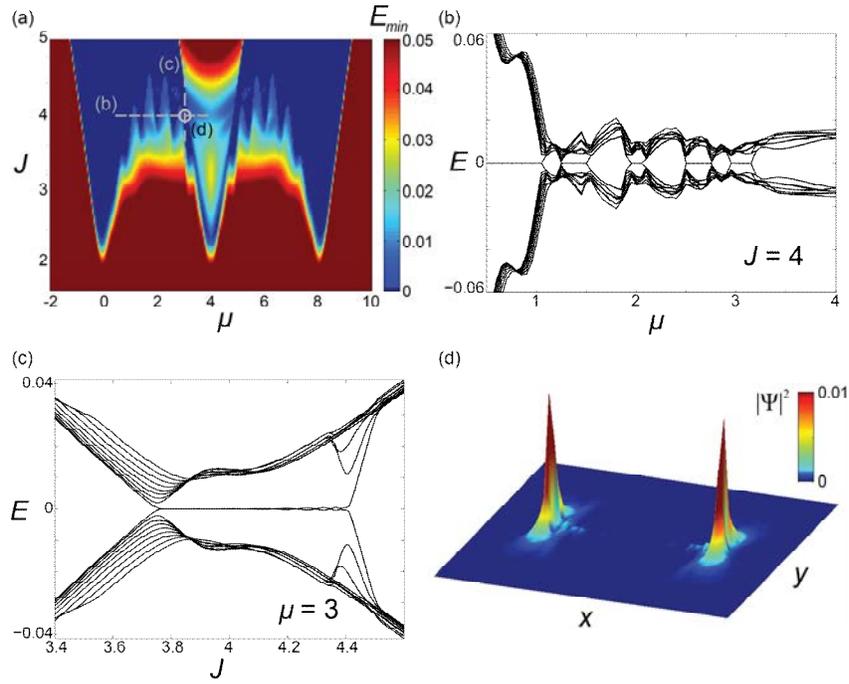

**Fig. S3**

(a) Phase diagram plotted in terms of the lowest energy of the eigenstates in a finite-size ferromagnetic chain embedded in a 2D superconductor lattice (Fig. S2a) (100 sites in a 120×21 lattice), and two line cuts showing the low-energy spectrum along lines with (b) constant J and (c) constant $\mu$. The topologically nontrival phase in (a) are shaded blue, while the red regions are trivial. The wavefunction of the lowest energy eigenstate corresponding to the crosspoint of the two line cuts is plotted in (d). Both lines and the crosspoint are indicated in (a).

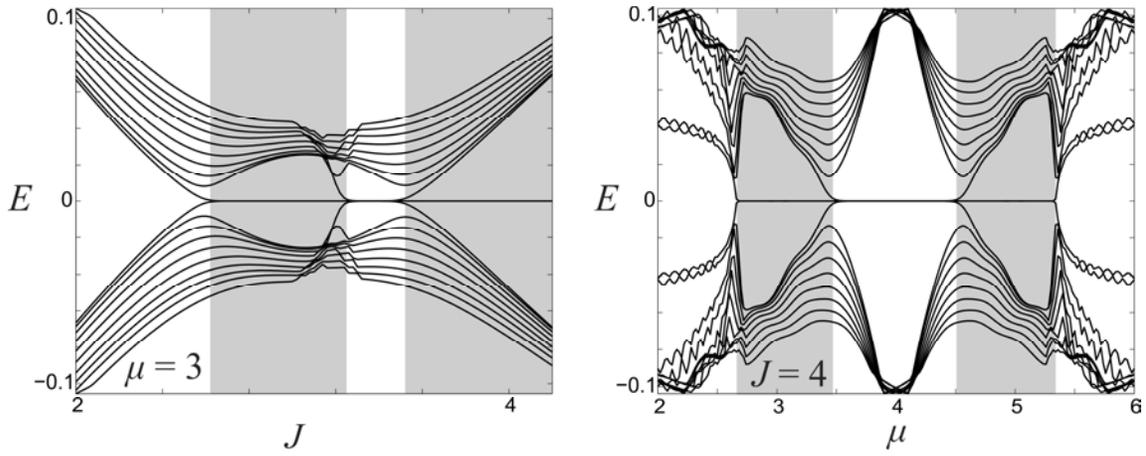

**Fig. S4**

Low-energy spectrum of two coupled finite-size Shiba chains (50×2 in an 80×32 lattice; see Fig. S2 (b)). The topologically nontrivial regions have been highlighted by a gray background.

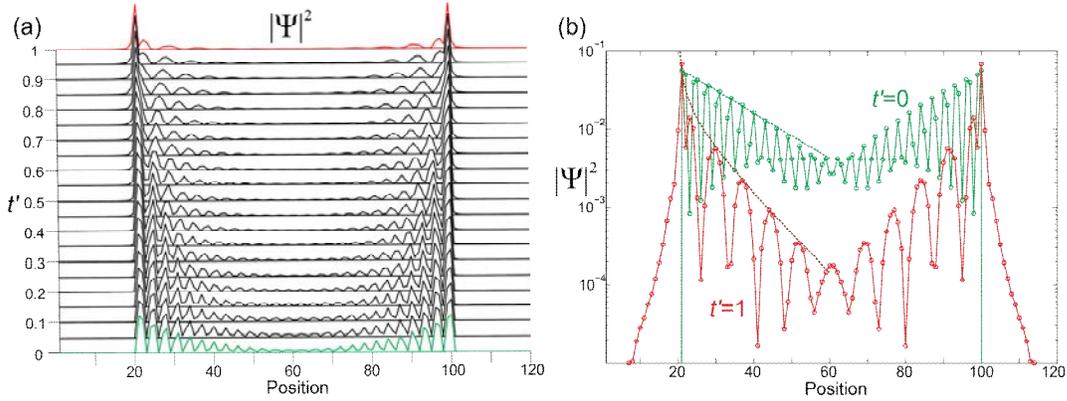

**Fig. S5**

Wavefunctions of the (nearly) zero-energy eigenstates, along the line of sites where the Shiba chain resides. Sharp drops in the amplitude of the wavefunctions are near the ends of the chain.. The evolution of the wavefunctions (in linear scale) as $t'$ is tuned from 0 to 1 is shown in (a). A comparison of the wavefunctions (in logarithm scale) between the two cases with $t'=0$ and $t'=1$ is shown in (b). Also shown in (b) are fittings (dotted lines) by exponential functions without (for $t'=0$) and with (for $t'=1$) a powerlaw prefactor.

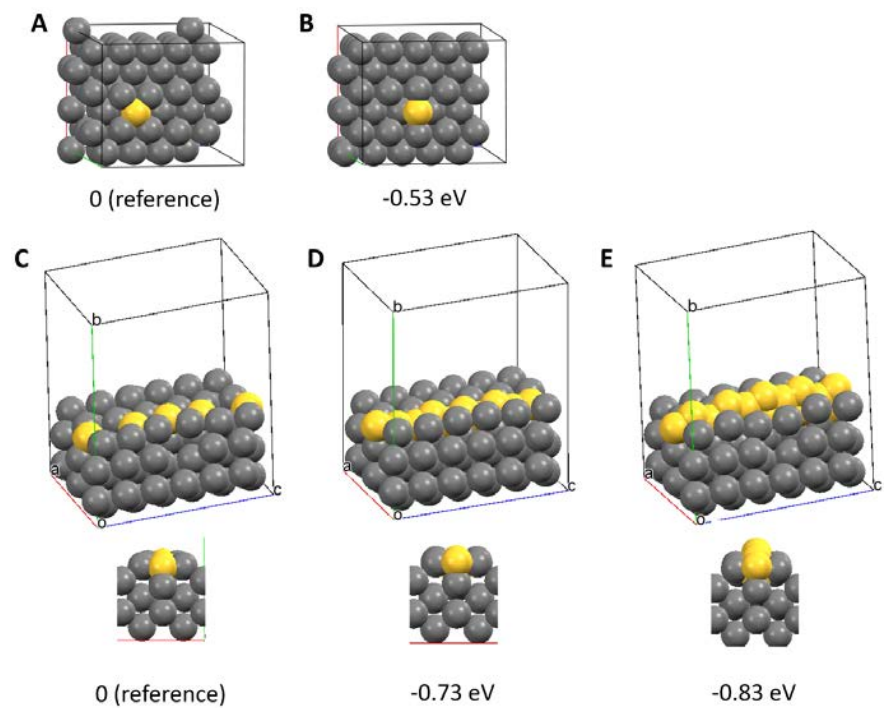

**Fig. S6**
(A and B) Top views of two stable surface adsorption sites for a single Fe atom on Pb(110). (C-E): Fe chains on Pb (110) with (C) 1-, (D) 2-, and (E) 3-layers of Fe atoms. The lower panels show side views along the chain direction. The numbers in each panel are relative formation energies per Fe atom.

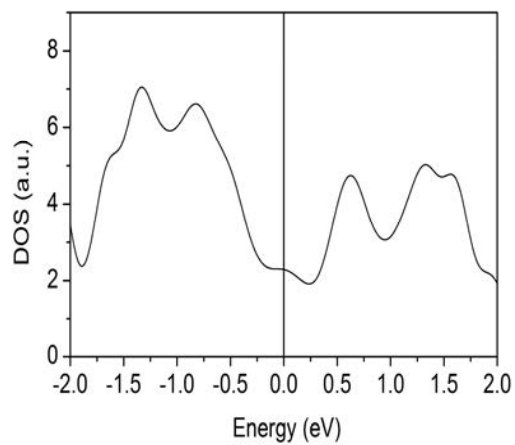

**Fig. S7**
DOS of the zigzag triple Fe chain embedded in a unstrained Pb surface shows a double peak structure seen experimentally and in the tight-binding model.

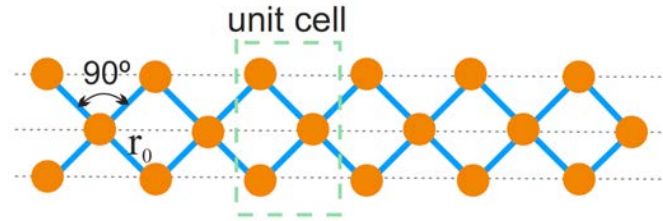

**Fig. S8**
Triple ziagzag chain structure used in the tight binding calculation

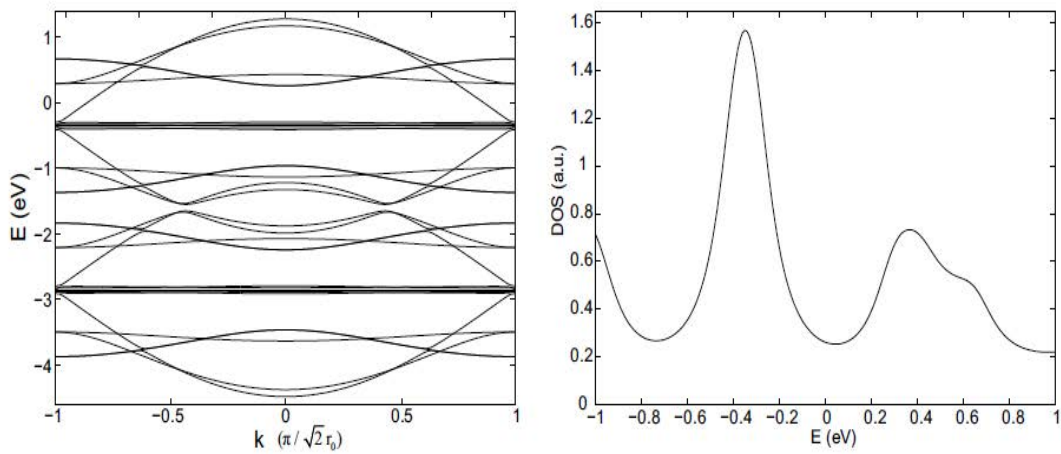

**Fig. S9**
Tight binding band structure of the Fe chain shown in Fig. S8 and its DOS as function of energy.

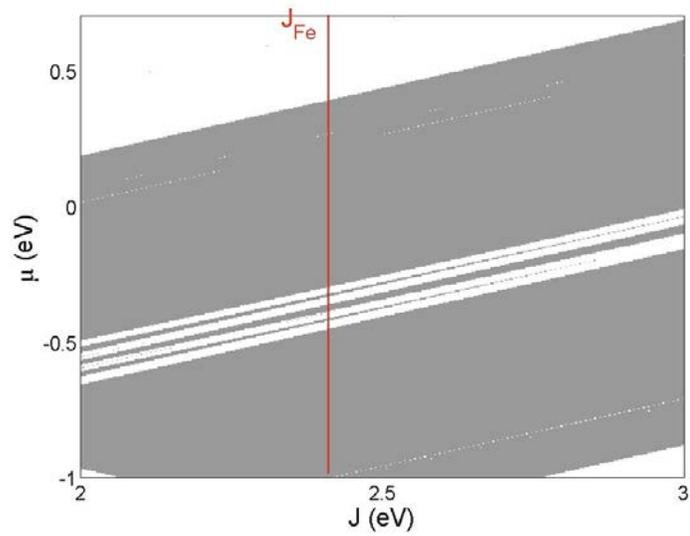

**Fig. S10**

Regime for trivial (white) and topological superconducting phases (gray) are identified for the band structure shown in Fig. S9 as a function of exchange interaction $J$ and chemical potential $\mu$, with the value for Fe chain based on DFT calculations marked as the red line.

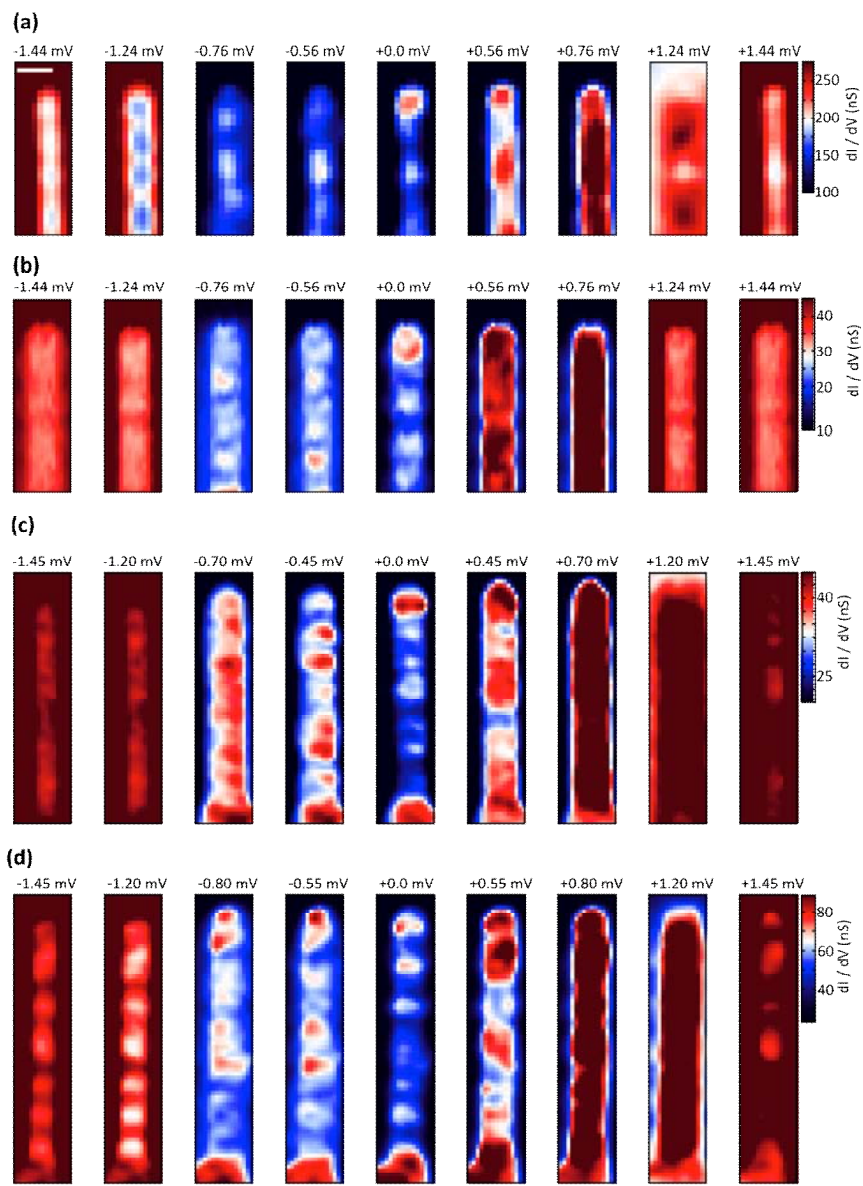

**Fig. S11.**

(a-d) spatially resolved conductance maps for four different atomic chains. The middle panel in each case shows the conductance map at zero bias.

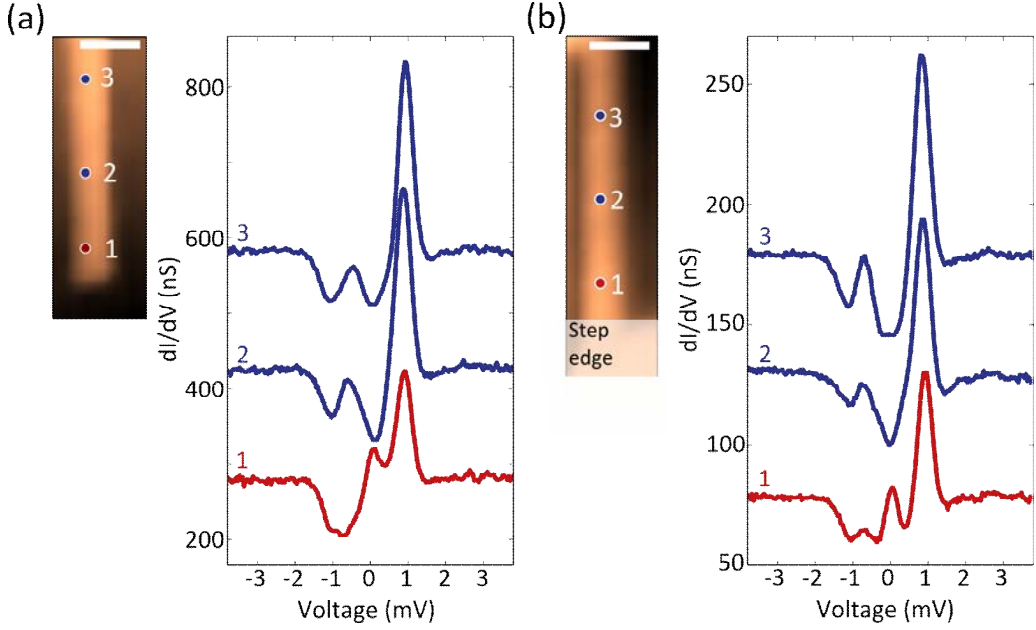

**Fig. S12.**

Point spectrum on two additional atomic chains. (a) Left: topography of the chain corresponding to chain 1 in Fig. S10. Panel (b) corresponds to the wire shown in Fig. 4D of the main text. Scale bars correspond to 20 Å.

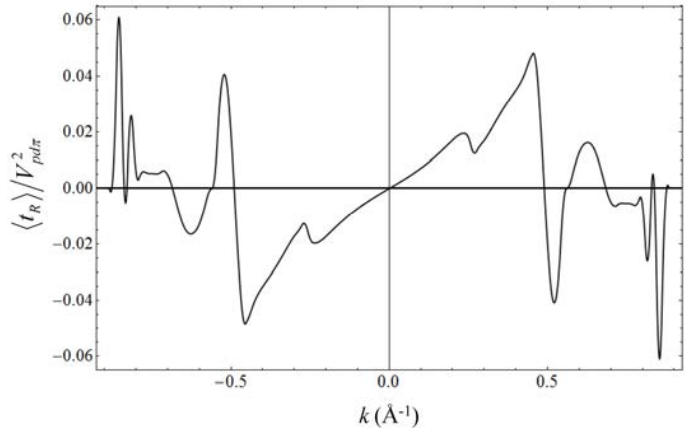

**Fig. S13.**

Effective Rashba spin-orbit coupling in Iron atomic chains as a function of momentum $k$.

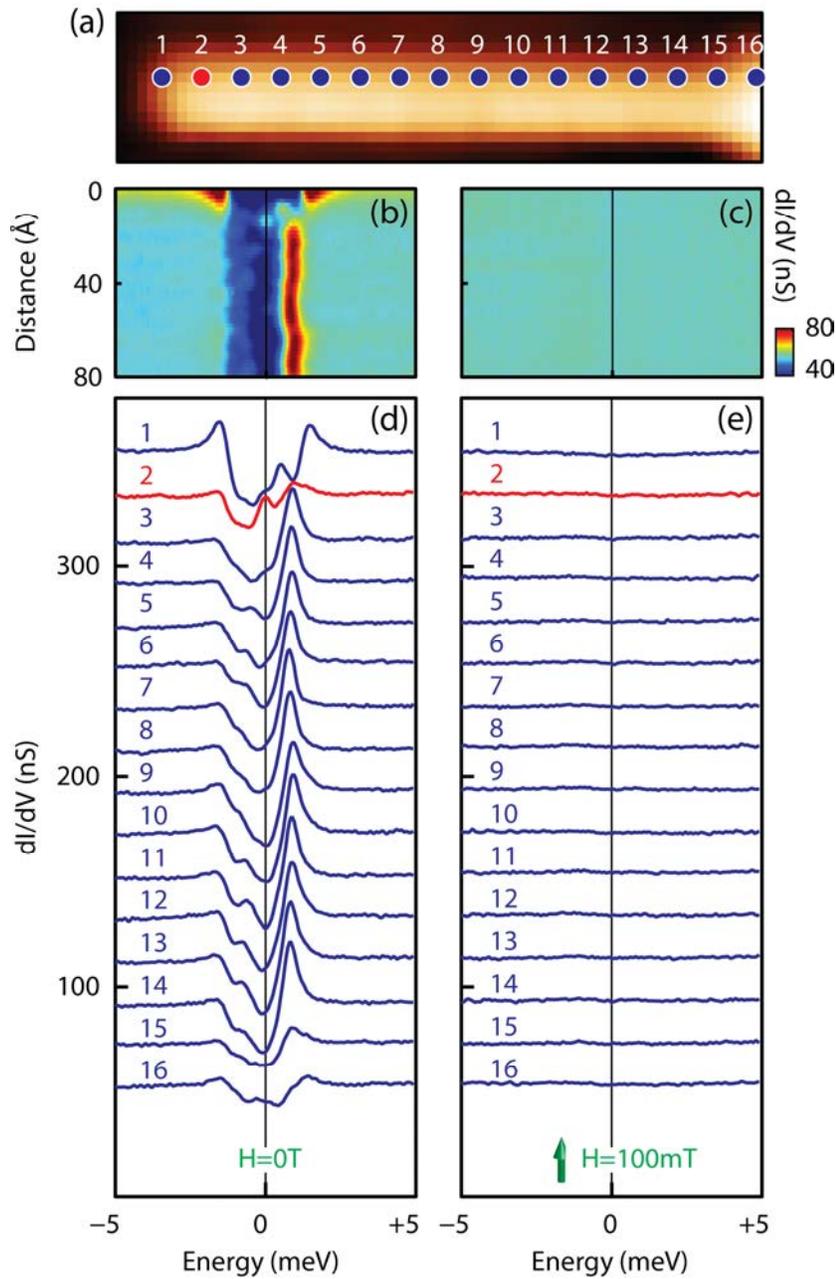

**Fig. S14.**
Spectroscopy of the atomic chain at zero and at non-zero magnetic field. (a) Topography (size of the image is 30 Å × 130 Å) and the points at which point spectra is taken in (d) and (e). (b) Differential conductance map plotted as function of position and energy, with zero distance corresponding to edge of topograph in (a). The ZBP feature is distinct from Shiba feature at all locations including at the ends. (c) same as (b) in $H$=100mT (d) Point spectra at $H = 0$ mT when substrate is in superconducting state. (e) Point spectra at $H = $ 100mT when substrate is in normal state. In the normal state the conductance is constant without any reminiscent features. Spectra in (b) and (c) are offset by 20nS for clarity.

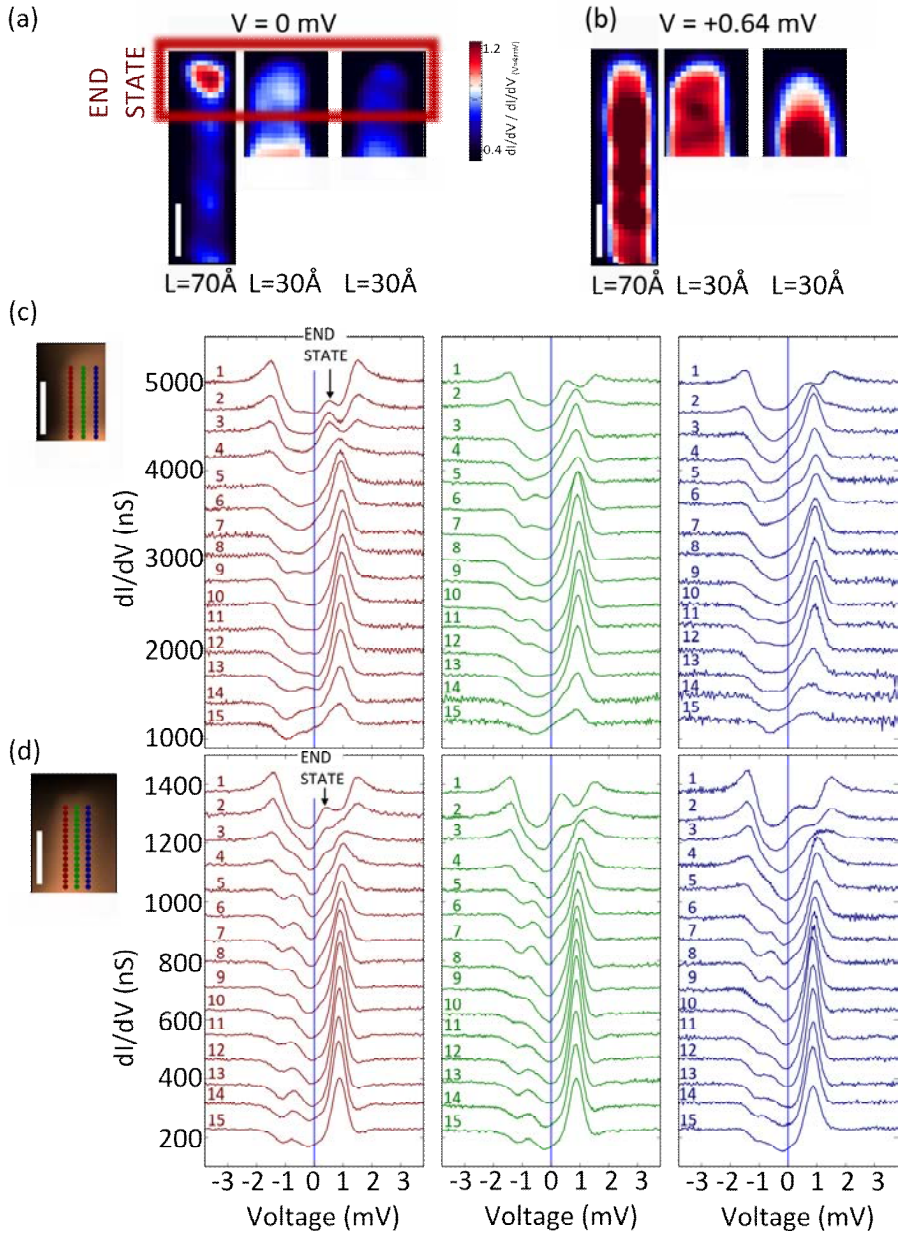

**Fig. S15.**
(a) and (b) Normalized spatial conductance map ($dI/dV$ / $dI/dV$ (4mV)) for $V = 0$ mV and $V = +0.64$ mV. Note the suppression of the end state (for $V = 0$ mV) in the case of short chains. Overall the normalized conductance is approximately equal (panel (b)). (c) and (d) show a spectral map (of dimension 15×3) on the short chain (insets show topography and position at which point spectra was taken). State at the end appears to be at $V$=300±50 μeV.

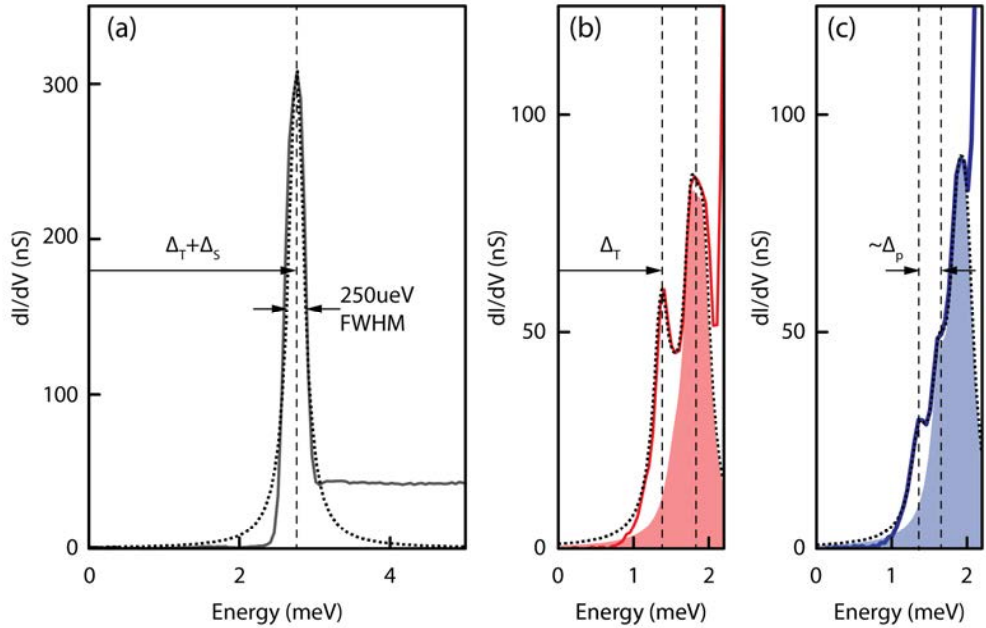

**Fig. S16.**
Line width analysis of the data in Figure 5 of the main text with a superconducting tip. (a) Superconducting tip tunneling into Pb(110) substrate, the experimental data in solid curve are compare to a broadened Lorentzian (dashed line). (b) Tunneling with the same tip on the end of the chain and modeling the thermal background from higher energy features (shaded areas) together with broadened ZBP as described in Section 9 to model the spectra. (c) Tunneling with the superconducting tip in the middle of the substrate and similar analysis as in (b).

| Parameters | Value (eV) |
|---|---|
| $V_{dd\sigma}$ | -0.6702 |
| $V_{dd\pi}$ | 0.5760 |
| $V_{dd\delta}$ | -0.1445 |

**Table S1**

Slater-Koster tight-binding model parameters for Fe (in eV). The hopping integral values are for the nearest-neighbor distance of bulk Fe(bcc), which is 2.383 Å.